\documentclass[prd,twocolumn,superscriptaddress,amsfonts,amssymb,amsmath,showpacs]{revtex4-2}
\usepackage{bm}
\usepackage{amsfonts}
\usepackage{latexsym}
\usepackage[latin1]{inputenc}
\usepackage{graphicx}
\usepackage{amsmath}
\usepackage{palatino}
\usepackage{mathpazo}
\usepackage{textcomp}
\usepackage{float}
\usepackage{booktabs}
\usepackage{dcolumn}
\usepackage{booktabs}
\usepackage{multirow}
\usepackage{hyperref}
\hypersetup{colorlinks=true,urlcolor=magenta,citecolor=red}
\usepackage{amsmath}
\usepackage{xcolor}
\usepackage{orcidlink}
\usepackage[caption=false]{subfig}
\usepackage{commath}
\def\jnl@style{\it}
\def\aaref@jnl#1{{\jnl@style#1}}

\def\aaref@jnl#1{{\jnl@style#1}}

\def\aj{\aaref@jnl{AJ}}                   
\def\apj{\aaref@jnl{ApJ}}                 
\def\apjl{\aaref@jnl{ApJ}}                
\def\apjs{\aaref@jnl{ApJS}}               
\def\apss{\aaref@jnl{Ap\&SS}}             
\def\aap{\aaref@jnl{A\&A}}                
\def\aapr{\aaref@jnl{A\&A~Rev.}}          
\def\aaps{\aaref@jnl{A\&AS}}              
\def\mnras{\aaref@jnl{Mon.~Not.~Roy.~Astron.~Soc.}}             
\def\prd{\aaref@jnl{Phys.~Rev.~D}}        
\def\prc{\aaref@jnl{Phys.~Rev.~C}}  
\def\prl{\aaref@jnl{Phys.~Rev.~Lett.}}    
\def\qjras{\aaref@jnl{QJRAS}}             
\def\skytel{\aaref@jnl{S\&T}}             
\def\ssr{\aaref@jnl{Space~Sci.~Rev.}}     
\def\zap{\aaref@jnl{ZAp}}                 
\def\nat{\aaref@jnl{Nature}}              
\def\aplett{\aaref@jnl{Astrophys.~Lett.}} 
\def\apspr{\aaref@jnl{Astrophys.~Space~Phys.~Res.}} 
\def\physrep{\aaref@jnl{Phys.~Rep.}}      
\def\physscr{\aaref@jnl{Phys.~Scr}}       
\def\commat{\aaref@jnl{Comm.~Math.~Phys.}}              
\def\science{\aaref@jnl{Science}}               
\def\cqg{\aaref@jnl{Classical Quant.~Grav.}}            
\def\jpcs{\aaref@jnl{JPCS}}                                     
\def\ijmpd{\aaref@jnl{Int.~J.~Mod.~Phys.~D}}                    
\def\grg{\aaref@jnl{Gen.~Relat.~Gravit.}}               
\def\rpp{\aaref@jnl{Rep.~Prog.~Phys.}}          
\def\npa{\aaref@jnl{Nucl.~Phys.~A}}        
\def\lrr{\aaref@jnl{Living Rev.~Rel.}}                   
\def\jcap{\aaref@jnl{J.~Cosmology Astropart.~Phys.}}    
\def\rmp{\aaref@jnl{Rev.~Mod.~Phys.}}   
\def\epjc{\aaref@jnl{Eur.~Phys.~J.~C}}

\allowdisplaybreaks[1]
\renewcommand{\arraystretch}{1.1}
\begin{document}

\color{black}       
\title{\bf Evolutionary behaviour of cosmological parameters with dynamical system analysis in $f(Q,T)$ gravity}

\author{Laxmipriya Pati\orcidlink{0000-0003-0828-192X}}
\email{lpriyapati1995@gmail.com}
\affiliation{Department of Mathematics,
Birla Institute of Technology and Science-Pilani, Hyderabad Campus,
Hyderabad-500078, India.} 

\author{S. A. Narawade\orcidlink{0000-0002-8739-7412}}
\email{shubhamn2616@gmail.com}
\affiliation{Department of Mathematics,
Birla Institute of Technology and Science-Pilani, Hyderabad Campus,
Hyderabad-500078, India.}

\author{S.K. Tripathy\orcidlink{0000-0001-5154-2297}}
\email{tripathy\_sunil@rediffmail.com }
\affiliation{Department of Physics, Indira Gandhi Institute of Technology, Sarang, Dhenkanal, Odisha-759146, India.}

\author{B. Mishra\orcidlink{0000-0001-5527-3565} }
\email{bivu@hyderabad.bits-pilani.ac.in }
\affiliation{Department of Mathematics,
Birla Institute of Technology and Science-Pilani, Hyderabad Campus,
Hyderabad-500078, India.}

\date{\today}
\begin{abstract}
We have investigated the accelerating behaviour of the universe in $f(Q,T)$ gravity in an isotropic and homogeneous space-time. We have initially derive the dynamical parameters in the general form of $f(Q,T)=\alpha Q^m+\beta T$ [Xu et al., Eur. Phys. J. C, \textbf{79}, 708 (2019)] and then split it into two cases (i) one with $m=1$ and the (ii) other with $\beta=0$. In the first case, it reduces to the linear form of the functional $f(Q,T)$ and second case leads to the higher power of the nonmetricity $Q$. In an assumed form of the hyperbolic scale factor, the models are constructed and its evolutionary behaviours are studied. The geometrical parameters as well the equation of state parameter are obtained and found to be in the preferred range of the cosmological observations. Marginal variation has been noticed in the behaviour of $\omega$ and $\omega_{eff}$ at present time. The violation of strong energy conditions in both the cases are shown. The dynamical system analysis for the models has been performed. 
\end{abstract}

\maketitle
\textbf{Keywords}:  Symmetric teleparallel gravity, Dynamical system analysis, Equation of state.

\section{Introduction}
The recent observational advances in cosmology have provided strong evidence that recently our Universe did enter in an accelerated expansion phase \cite{Riess1998, Perlmutter1999, Bernardis2000, Hanany2000, Ade2016, Akrami2018, Aghanim2016}. As a result of these observations, standard general relativity (GR) may not be adequate to explain gravitational phenomena on galactic and cosmological scales, despite its achievements and remarkable success at the Solar System scale. Due to these limitations, standard general relativity is not sufficient to explain the two fundamental problems that current cosmology faces: the dark matter problem as well as the dark energy problem. Many classical approaches have been proposed in order to explain the observational results of cosmology. Recently another attempt has been taken to address the late time cosmic acceleration issue by proposing a new gravitational theory, the $f(Q,T)$ theory of gravity. Xu et al. \cite{Xu19} have proposed this modified theory of gravity by extending the symmetric teleparallel gravity. In $f(Q,T)$ gravity, $Q$ be the nonmetricity and $T$ be the trace of the energy momentum tensor. We shall give here a brief discussion on the development of this modified theory.

It can be inferred that GR can be represented geometrically at least with curvature representation and teleparallel representation. The torsion and curvature vanishes respectively in the curvature and teleparallel representation whereas the nonmetricity vanishes in both the approaches. Hence another approach is on the description of non-vanishing of the basic geometrical variable, the nonmetricity. This approach is known as the symmetric teleparallel gravity proposed by Nester and Yo \cite{Nester99}. Most recently it has been developed as the $f(Q)$ gravity by Jimenez et al. \cite{Jimenez18}. Latorre et al. \cite{Latorre18} have investigated that the nonmetricity produces observable effects in the quantum fields. Conroy and Koivisto \cite{Conroy18} have given a  note on the spectrum of symmetric teleparallel gravity.  Soudi et al. \cite{Soudi19} have studied the gravitational waves in $f(Q)$ gravity and obtained the same speed and polarization as in GR. Lu et al. \cite{Lu19} in $f(Q)$ gravity, have indicated that the role of dark energy can be played by the geometry itself. Bajardi et al. \cite{Bajardi20} have investigated the bouncing cosmology in symmetric teleparallel gravity. It has been claimed that the nonmetricity gravity can challenge the $\Lambda$CDM \cite{Anagnostopoulos21}. Narawade et al. \cite{Narawade22} have shown the stability of the $f(Q)$ gravity model with dynamical system analysis. For the different choice of $f(Q)$, several perturbative corrections to the Schwarzschild solution have also been shown \cite{Ambrosio22}. The cosmological model of the Universe has been presented in $f(Q)$ gravity and the parameters have been constrained from the cosmological data sets in Ref.\cite{Narawade2022}.

Xu et al. \cite{Xu19} extended $f(Q)$ gravity by introducing the non-minimal coupling between the nonmetricity and the trace of energy momentum tensor $T$. The Lagrangian density of the  gravitational field would be described with respect to $Q$ and $T$ in the form $L=f(Q,T)$. The motivation behind this action is to study the cosmological implications such as, to describe the decelerating and accelerating evolutionary phase of the Universe. Some $f(Q,T)$ gravity cosmological models are available in some recent literature. Xu et al. \cite{Xu20} have given the Weyl type $f(Q,T)$ gravity and its cosmological implications. Zia et al. \cite{Zia21} have presented transit cosmological model aligning with the observational value of the deceleration parameter in $f(Q,T)$ gravity. Pati et al. \cite{Pati21a} have obtained quintessence model in the context of hybrid scale factor. Agrawal et al. \cite{Agrawal21} have shown the non-singular matter bouncing scenario with the violation of null and strong energy conditions. Najera and Fajardo \cite{Najera21} have tested five $f(Q, T)$ models and have shown that with certain values of the parameters, the models reduce to $\Lambda$CDM model. Godani and Samanta \cite{Godani21} have studied the FRW cosmology in $f(Q,T)$ gravity and compared the results with that of $\Lambda$CDM model. Pati et al. \cite{Pati21b} have presented the non-occurrence of singularity in the form of rip cosmology in $f(Q,T)$ gravity. Some more works on $f(Q,T)$ gravity are available in the literature \cite{Yang21, Iosifidis21, Najera22, Pradhan21, Shiravand22, Sokoliuk22}. Recently, several cosmological models area available in the literature on the dynamical system to reveal the evolutionary behaviour of the dark energy models \cite{Khyllep21, Khyllep22, Agrawal22, Duchaniya22, Duchaniya22a, Duchaniya23, Bonanno12} in the modified theories of gravity. This motivates us to study the cosmological aspects of the models through the dynamical system analysis in symmetric teleparallel gravity.

The paper is organised as follows: in Sec. \ref{Sec.II}, the $f(Q,T)$ gravity has been discussed and the field equations are derived in FLRW space-time. For the general case of $f(Q,T)=\alpha Q^m+\beta T$, the dynamical parameters and energy conditions are given in Sec. \ref{Sec.III}. In Sec. \ref{Sec.IV}, we have presented two cases by considering (i) $m=1$, which reduces to linear case of $f(Q,T)$ and (ii) $\beta=0$ that reduces to the $f(Q)$ gravity. The cosmological models for both the cases are constructed with the hyperbolic Hubble parameter. In Sec. \ref{Sec.V} the dynamical system of the models has been perfomed. Finally the results and conclusions are given in Sec. \ref{Sec.VI}.  

\section{Overview of the $f(Q,T)$ gravity and derivation of the dynamical parameters} \label{Sec.II}
The action of $f(Q,T)$ gravity can be given as \cite{Xu19},

\begin{equation}\label{eq:1}
S=\int\left[\dfrac{1}{16\pi}f(Q,T)d^{4}x\sqrt{-g}+\mathcal{L}_{m} d^{4}x\sqrt{-g}\right]
\end{equation}
where $Q$ and $T$ in the functional $f(Q,T)$ respectively represents the nonmetricity and the trace of the energy momentum tensor $T_{\mu\nu}$. $\mathcal{L}_m$ denotes the matter Lagrangian and $g=det(g_{\mu\nu})$ be the determinant of the metric tensor $g_{\mu\nu}$. In differential geometry, affine connections can be decomposed into three parts,
\begin{equation}\label{eq:2}
\Gamma^{\lambda}{}_{~\mu\nu} = \{^{\lambda}{}_{~\mu \nu}\} + K^{\lambda}{}_{~\mu \nu} + L^{\lambda}{}_{~\mu\nu}
\end{equation}
For arbitrary connection, the nonmetricity scalar is defined as,
\begin{widetext}
\begin{equation}\label{eq:3}
    Q=-Q_{\mu\nu\rho}P^{\mu\nu\rho}=\frac{1}{4}Q_{\mu\nu\rho}Q^{\mu\nu\rho}-\frac{1}{2}Q_{\mu\nu\rho}Q^{\mu\nu\rho}-\frac{1}{4}Q_{\rho\mu}{}^{\mu}Q^{\rho\nu}{}_{\nu}+\frac{1}{2}Q^{\mu}{}_{\mu\rho}Q^{\rho\nu}{}_{\nu}
\end{equation}
Here $P^{\mu\nu\rho}$ is called nonmetricity conjugate, which is defined as,\\
\begin{equation}\label{eq:4}
p^{\rho \mu\nu}=-\frac{1}{4}Q^{\rho \mu\nu}+\frac{1}{2}Q^{(\mu\nu)\rho}+\frac{1}{4}g^{\mu\nu}(Q^{\rho\sigma}{}_{\sigma}-Q_{\sigma}{}^{\sigma\rho})-\frac{1}{4}g^{\rho(\mu}Q^{\nu)\sigma}{}_{\sigma},
\end{equation}
\end{widetext}
The nonmetricity scalar is related to the ricci scalar of the Levi-Civita connection through a boundary term,\\
\begin{equation}\label{eq:5}
    \overset{LC}{R}=-\overset{STP}{Q}-\overset{LC}{\nabla}_{\mu}({\overset{STP}{Q^{\mu\nu}}}{}_{\nu}-\overset{STP}{Q}_{\nu}{}^{\nu\mu}).
\end{equation}
In \eqref{eq:2} the Levi-Civita connection of the metric tensor $g_{\mu \nu}$ is,  $\{^{\lambda}{}_{~\mu\nu}\} \equiv \frac{1}{2} g^{\lambda \alpha}\left({\partial}_{\mu} g_{\alpha \nu}+{\partial}_{\nu} g_{\alpha \mu}-{\partial}_{\alpha} g_{\mu \nu}\right)$, The contortion is defined as $K^{\lambda}{}_{~\mu \nu}\equiv \frac{1}{2}\left( T^{\lambda}{}_{~\mu \nu}+T_{\mu}{}^{\lambda}{}_{\nu}+T_{\nu}{}^{\lambda}{}_{\mu}\right)$, where the torsion is represented by $T^{\lambda}{}_{\mu\nu}\equiv \Gamma ^{\lambda}{}_{\mu\nu}- \Gamma ^{\lambda}{}_{\nu\mu}$. As we know, the nonmetricity tensor consists of, $Q_{\rho \mu\nu}\equiv \nabla_{\rho}g_{\mu \nu}=\partial_{\rho}g_{\mu\nu}-\Gamma^{\beta}{}_{\rho \mu}g_{\beta \nu}-\Gamma^{\beta}{}_{\rho \nu}g_{\beta \mu}$. We are considering in Symmetric teleparallel theory where curvature and torsion is imposed to be zero. Due to this reason contortion should vanishes. Also we have fixing the gauge namely coincident gauge, then we got nonmetricity tensor is equal to the partial derivative of the fundamental metric. Mathematically, $Q_{\rho \mu\nu}=\partial_{\rho}g_{\mu\nu}$. In addition to this, Due to this gauge choice from \eqref{eq:2} we obtained the disformation tensor is equals to negative sign of Levi-Civita connection. Mathematically denoted as, $L^{k}{}_{l\gamma}\equiv-\frac{1}{2}g^{k\lambda}(\partial_{\gamma}g_{l\lambda}+\partial_{l}g_{\lambda \gamma}-\partial_{\lambda}g_{l\gamma})$. After fixing the gauge the nonmetricity scalar becomes,
\begin{equation}\label{eq:6}
   Q\equiv -g^{\mu \nu}( L^k_{~l\mu}L^l_{~\nu k}-L^k_{~lk}L^l_{~\mu \nu}),
\end{equation}
By varying \eqref{eq:1} the general field equation becomes \cite{Xu19},

\begin{widetext}
\begin{equation}\label{eq:7}
-\frac{2}{\sqrt{-g}}\bigtriangledown_{k}(F\sqrt{-g}p^{k}{}_ {\mu \nu})-\frac{1}{2}fg_{\mu \nu}-F(p_{\mu kl} Q_{\nu}{}^ {kl}-2Q^{kl}{}_{\mu} p_{kl\nu})=8 \pi T_{\mu \nu}\left(1-\kappa\right)-8\pi\kappa \Theta_{\mu \nu}.
\end{equation}
\end{widetext}
For brevity, we represent $f\equiv f(Q,T)$ and  $F=\frac{\partial f}{\partial Q}$, $8\pi\kappa=\frac{\partial f}{\partial T}$. Further the super potential of the model, the energy momentum tensor and the trace of the nonmetricity can be obtained respectively as,

\begin{eqnarray}
p^{k}{}_{\mu \nu}&=&-\frac{1}{2}L^{k}{}_{\mu \nu}+\frac{1}{4}(Q^{k}-\tilde{Q}^{k})g_{\mu \nu}-\frac{1}{4}\delta^{k}{}_{(\mu}Q_{\nu)}, \nonumber \\
T_{\mu \nu}&=&\frac{-2}{\sqrt{-g}} \frac{\delta(\sqrt{-g}L_{m})}{\delta g^{\mu \nu}};~~~~~ \Theta_{\mu \nu}=g^{kl}\frac{\delta T_{kl}}{\delta g^{\mu \nu}}, \nonumber\\
Q_{k}&=&Q_{k}^{\;\;\mu}\;_{\mu},~~~~~~~ \tilde{Q}_{k}=Q^{\mu}\;_{k\mu}.\label{eq:8}
\end{eqnarray}

We wish to study the cosmological model of the Universe in $f(Q,T)$ theory of gravity at the background of an isotropic and homogeneous FLRW space-time considered in the form, 
\begin{eqnarray}\label{eq:9}
ds^{2}=-N^{2}(t)dt^{2}+a^{2}(t)(dx^{2}+dy^{2}+dz^{2}),
\end{eqnarray}
where the lapse function $N(t)$ and scale factor $a(t)$ are the function of cosmic time. Also, the dilation rate can be defined as, $\tilde{T}=\frac{\dot{N}(t)}{N(t)}$. In an FLRW space-time, the lapse function can be $N(t)=1$ and subsequently the dilation rate, $\tilde{T}=0$ and the nonmetricity reduces to $Q=6H^2$, where $H=\frac{\dot{a}}{a}$ is the Hubble parameter. We consider the energy momentum tensor that of a perfect fluid distribution  $T^{\mu}_{\nu}=diag(-\rho,p,p,p)$. Also, $\Theta^{\mu}_{\nu}=diag(2\rho+p,-p,-p,-p)$. So, the field equations of $f(Q,T)$ gravity \eqref{eq:7} in FLRW space-time can be obtained as,

\begin{eqnarray}
-16\pi p&=&f-12\frac{\chi^2}{F}-4\dot{\chi},\label{eq:10}   \\ 
16\pi\rho&=&f-12\frac{\chi^2}{F}-4\dot{\chi}\kappa_1,\label{eq:11} 
\end{eqnarray}
where $\chi=FH$ and $\kappa_1=\frac{\kappa}{1+\kappa}$.  Also, we have $\dot{\chi}=F\dot{H}+\dot{F}H$. From Eqs. \eqref{eq:10} and \eqref{eq:11}, the evolution equation for $\chi$ can be obtained as,
\begin{equation}
\dot{\chi}=4\pi\left(\rho+p\right)\left(1+\kappa\right). \label{eq:12}
\end{equation}
For a constant value of $F$, the above evolution equation reduces to the evolution equation for the Hubble parameter as,
\begin{equation}
\dot{H}=\frac{4\pi}{F}\left(1+\kappa\right)\left(\rho+p\right). \label{eq:13}
\end{equation}
Assuming a barotropic relationship $p=\omega\rho$, the above relation reduces to,
\begin{equation}
\rho=\frac{F}{4\pi\left(1+\kappa\right)(1+\omega)}\dot{H}. \label{eq:14}
\end{equation}
The equation of state (EoS) parameter  $\omega=\frac{p}{\rho}$ for the $f(Q,T)$ gravity theory may be obtained from Eqs. \eqref{eq:10} and \eqref{eq:11} as,
\begin{equation}
\omega=-1+\frac{4\dot{\chi}}{\left(1+\kappa\right)\left(f-12F H^2\right)-4\dot{\chi}\kappa}. \label{eq:15}
\end{equation}

The EoS parameter will decide the possibility of the accelerating Universe at least at the late times of the cosmic evolution.

The equivalent Friedmann equations for the present gravity theory may be written as, 
\begin{widetext}
\begin{eqnarray}
2\dot{H}+3H^2&=&\frac{1}{F}\left[\frac{f}{4}-2\dot{F}H+4\pi\left[(1+\kappa)\rho+(2+\kappa)p\right]\right]=-8\pi p_{eff},\label{eq:16} \\
3H^2&=& \frac{1}{F}\left[\frac{f}{4}-4\pi\left[(1+\kappa)\rho+\kappa p\right]\right]=8\pi \rho_{eff}.  \label{eq:17}
\end{eqnarray}
\end{widetext}
Obviously, the effective energy density $\rho_{eff}$ and the effective pressure $p_{eff}$ satisfy the conservation equation
\begin{equation}
\dot{\rho}_{eff}+3H\left(\rho_{eff}+p_{eff}\right)=0.\label{eq:18}
\end{equation}

Consequently, we may define an effective EoS parameter as
\begin{equation}
\omega_{eff}=-1-\frac{8 F\dot{H}}{f-16\pi\left[(1+\kappa)\rho+\kappa p\right]}. \label{eq:19}
\end{equation}

In order to investigate viable cosmological scenario in the framework of the above discussed $f(Q,T)$ gravity theory, it is required to consider certain assumed form of the functional $f(Q,T)$. In the seminal work, Xu et al. \cite{Xu19} have considered three different forms for $f(Q,T)$ such as (i) $f(Q,T)= \alpha Q+ \beta T$  (ii) $f(Q,T)= \alpha Q^{m}+ \beta T$ and (iii) $f(Q,T)=-\left(\alpha Q+ \beta T^2\right)$. Here $\alpha,~\beta$ and $m$ are constants. 

\section{Dynamical Parameters and Energy Conditions} \label{Sec.III}
In the present work, we will consider the functional $f(Q,T)=\alpha Q^m+\beta T$ to model the Universe. Also, we wish to obtain the corresponding models for the choice $m=1$ and $\beta=0$. It should be remarked here that, for $\beta=0$, the model reduces to $f(Q)$ gravity.
For this choice of the functional, we have  $F=\alpha mQ^{m-1}$, $\dot{F}=2(m-1)F\frac{\dot{H}}{H}$,  $\chi=\alpha mQ^{\left(m-1\right)}H$, $\dot{\chi}=F\dot{H}\left(2m-1\right)$ and $\beta=8\pi\kappa$. We may now have the dynamical parameters as,

\begin{eqnarray}
p &=& \frac{\left(2m-1\right)\alpha Q^m+2\dot{\chi}\left[2+\kappa-\kappa\kappa_1\right]}{4\pi\left[\left(2+\kappa\right)\left(2+3\kappa\right)-3\kappa^2\right]},\label{eq:20}\\
\rho &=& \frac{(1-2m)\alpha Q^m+2\dot{\chi}\left[3\kappa-\left(2+3\kappa\right)\kappa_1\right]}{4\pi\left[\left(2+\kappa\right)\left(2+3\kappa\right)-3\kappa^2\right]},\label{eq:21}\\
\omega&=&-1+\frac{4\dot{\chi}\left[(1+2\kappa)(1-\kappa_1)\right]}{(1-2m)\alpha Q^m+2\dot{\chi}\left[3\kappa-(2+3\kappa)\kappa_1\right]}\label{eq:22}.
\end{eqnarray}

We may express the above equations,  Eqns. \eqref{eq:20}-\eqref{eq:22} in the term of the Hubble parameter and the deceleration parameter $q=-1-\frac{\dot{H}}{H^2}$ as,
\begin{widetext}
\begin{eqnarray}
p &=&- \frac{2^m3^{(m-1)}(1-2m)\alpha H^{2m}\left[3-m(1+q)\left(2+\kappa-\kappa\kappa_1\right)\right]}{4\pi\left[\left(2+\kappa\right)\left(2+3\kappa\right)-3\kappa^2\right]},\label{eq:23}\\
\rho &=& \frac{2^m3^{(m-1)}(1-2m)\alpha H^{2m}\left[3+m(1+q)\left(3\kappa-\left(2+3\kappa\right)\kappa_1\right)\right]}{4\pi\left[\left(2+\kappa\right)\left(2+3\kappa\right)-3\kappa^2\right]},\label{eq:24}\\
\omega&=&-1+\frac{2m(1+q)\left[(1+2\kappa)(1-\kappa_1)\right]}{3+m(1+q)\left(3\kappa-\left(2+3\kappa\right)\kappa_1\right)}\label{eq:25}.
\end{eqnarray}
\end{widetext}

The effective EoS parameter \eqref{eq:19} becomes
\begin{equation}
\omega_{eff}=-1+\frac{2^{(m+2)}3^{m-1}\alpha mH^{2m}(1+q)}{6^{m}\alpha H^{2m}-16\pi\rho\left[(1+\kappa)+k\omega)\right]}. \label{eq:26}
\end{equation}
We say that the model is a cosmological constant ($\Lambda CDM$) whenever $\omega_{\rm eff} = -1$, a quintessence model when $-1 < \omega_{\rm eff} \leq- \frac{1}{3}$, and  a phantom model whenever $\omega_{\rm eff} < -1$. A number of cosmological analyses constrain the numerical value of the EoS parameter, including Supernovae Cosmology Project, in which $\omega_{-0.059}^{+0.055}$ \cite{Amanullah2010}; WAMP+CMB, $\omega_{\rm eff} = -1.079_{-0.089}^{+0.090}$ \cite{Hinshaw2013}; Plank 2018, $\omega_{\rm eff} = -1.03\pm 0.03$ \cite{Aghanim2020}.
Since the study of energy energy conditions is an important aspect of the gravitational theory, we have given below the energy conditions of the proposed problem as,
\begin{widetext}
\begin{eqnarray}
\rho+p&=&\frac{2^{m}3^{m-1}(1-2m)\alpha H^{2m}}{8\pi}[m(1+q)(1-\kappa_{1})], ~~~{\bf[NEC/WEC]} \label{eq:27} \\
\rho+3p&=&\frac{2^{m}3^{m-1}(1-2m)\alpha H^{2m}}{8\pi(1+2\kappa)}[-3+m(1+q)(3+3\kappa-\kappa_{1}-3\kappa\kappa_{1})], {\bf [SEC]}  \label{eq:28}\\
\rho-p&=&\frac{2^{m}3^{m-1}(1-2m)\alpha H^{2m}}{8\pi(1+2\kappa)}[3+m(1+q)(-1+\kappa-\kappa_{1}-\kappa\kappa_{1})],{\bf [DEC]},\label{eq:29}
\end{eqnarray}\label{EC}
\end{widetext}
where NEC, WEC, SEC and DEC respectively represent null, weak, strong and dominant energy conditions. All the dynamical parameters and the energy conditions are expressed in terms of Hubble and deceleration parameter. So, we consider in the subsequent section, a specific form of the Hubble parameter.  

\section{Models with Hyperbolic Scale Factor} \label{Sec.IV}
To analyse the evolutionary behaviour of the dynamical parameters, the Hubble parameter involved in  Eqns. \eqref{eq:23}-\eqref{eq:25} required to be expressed in terms of cosmic time or redshift, $1+z=\frac{1}{a(t)}$. In the literature several scale factors such as, de Sitter expansion \cite{Ronald65}, power law expansion \cite{Berman83}, hybrid scale factor \cite{Mishra15}, bouncing scale factor \cite{Agrawal21} and many more are introduced from time to time to address the astrophysical and cosmological issues of the Universe. In a similar approach, here we shall consider the Hubble parameter in such a manner that its corresponding scale factor would have a quadratic term in its exponent. The hyperbolic scale factor can be expressed in the form, 
\begin{equation}
a(t)\propto \left[\sinh\left(\frac{3}{2}\sqrt{\frac{\Lambda}{3}} t\right)\right]^\frac{2}{3},\label{eq:30}
\end{equation}
\begin{figure}[h]
\centering
\includegraphics[scale=0.5]{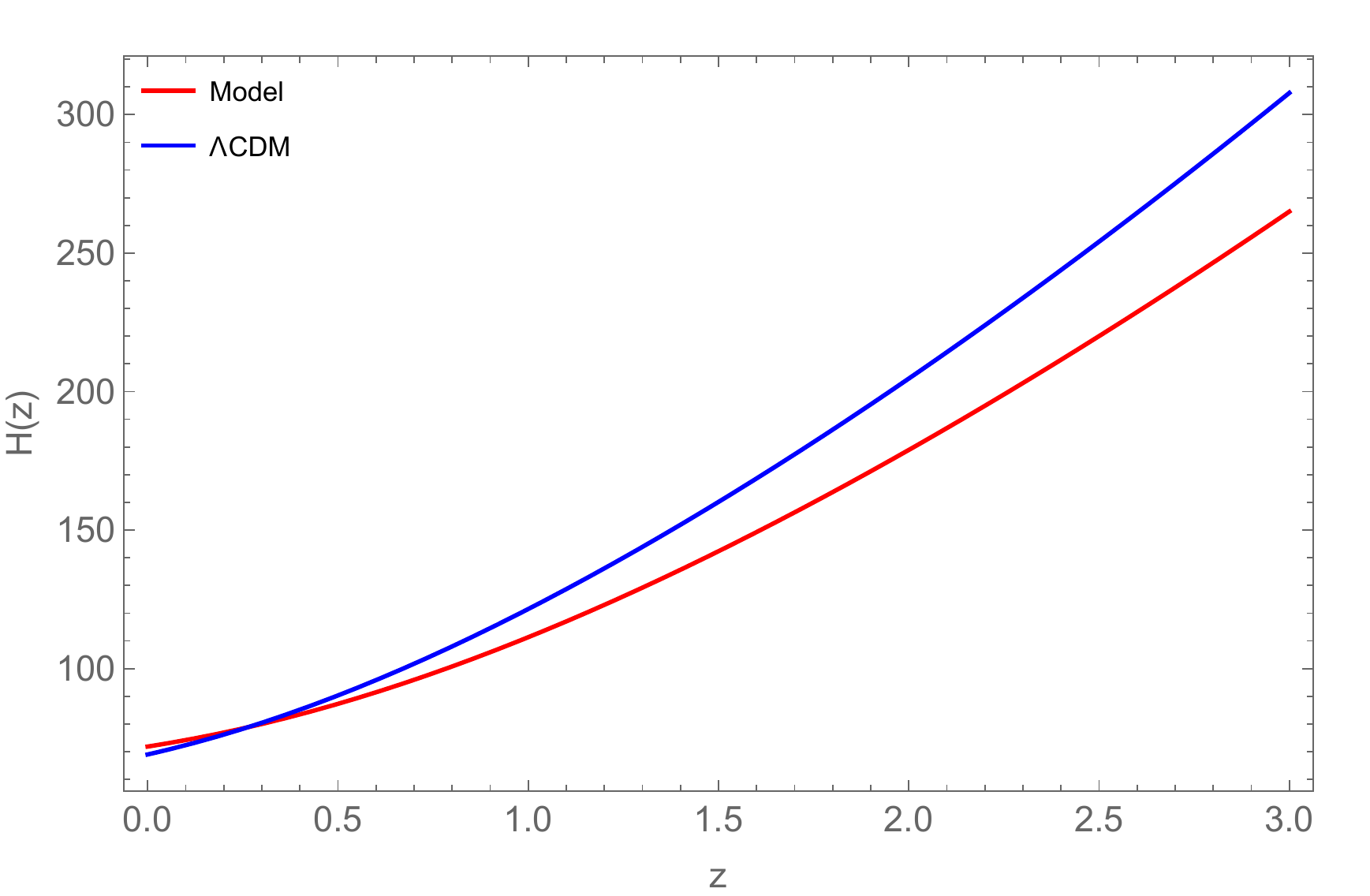}
\includegraphics[scale=0.5]{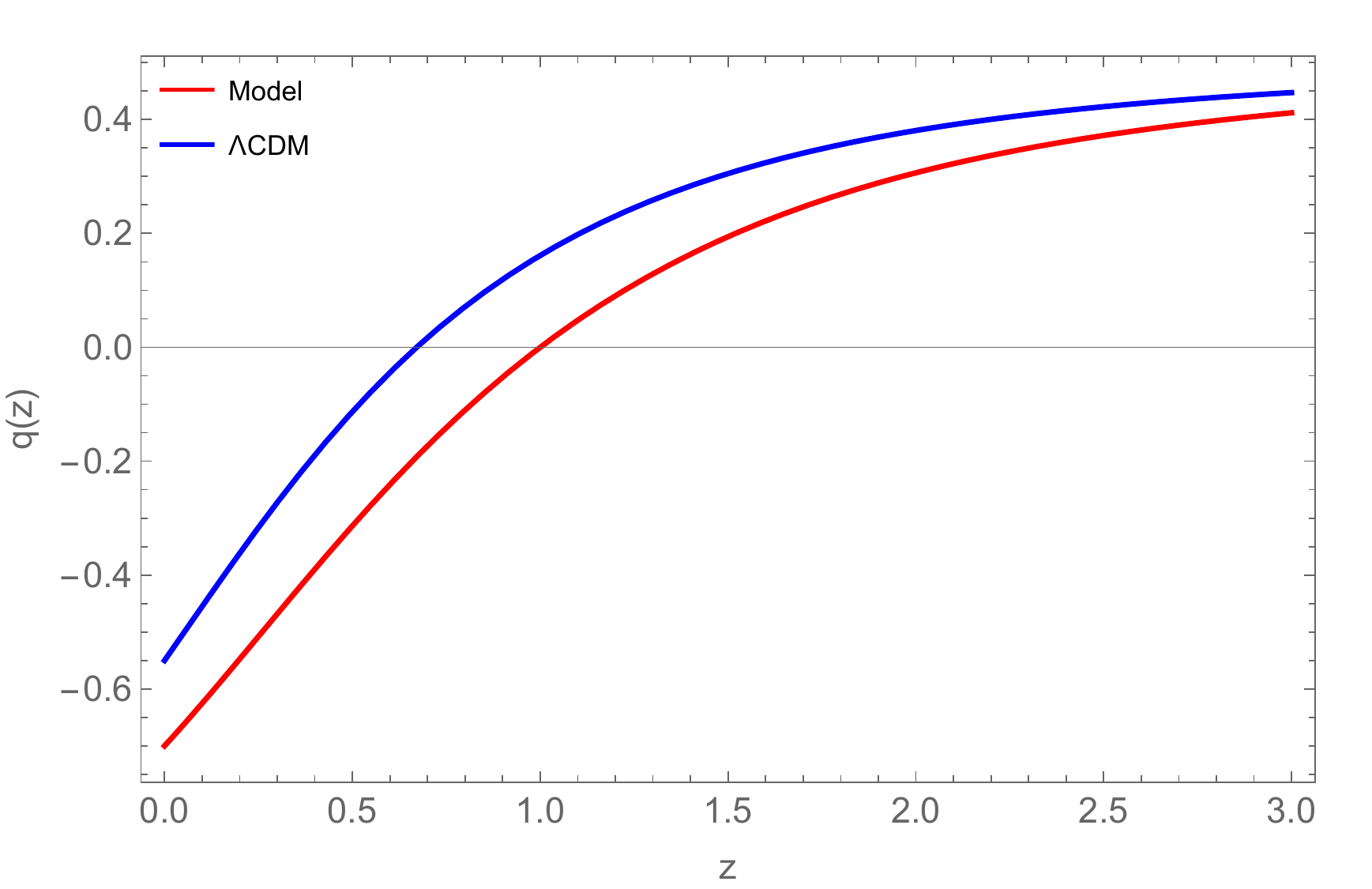}
\caption{Hubble parameter (Top panel) and deceleration parameter (Bottom panel) in redshift. The parameter scheme, $\Lambda=e^{3\pi}$.} \label{FIG--1}
\end{figure}
\begin{figure}[h]
\centering
\includegraphics[scale=0.5]{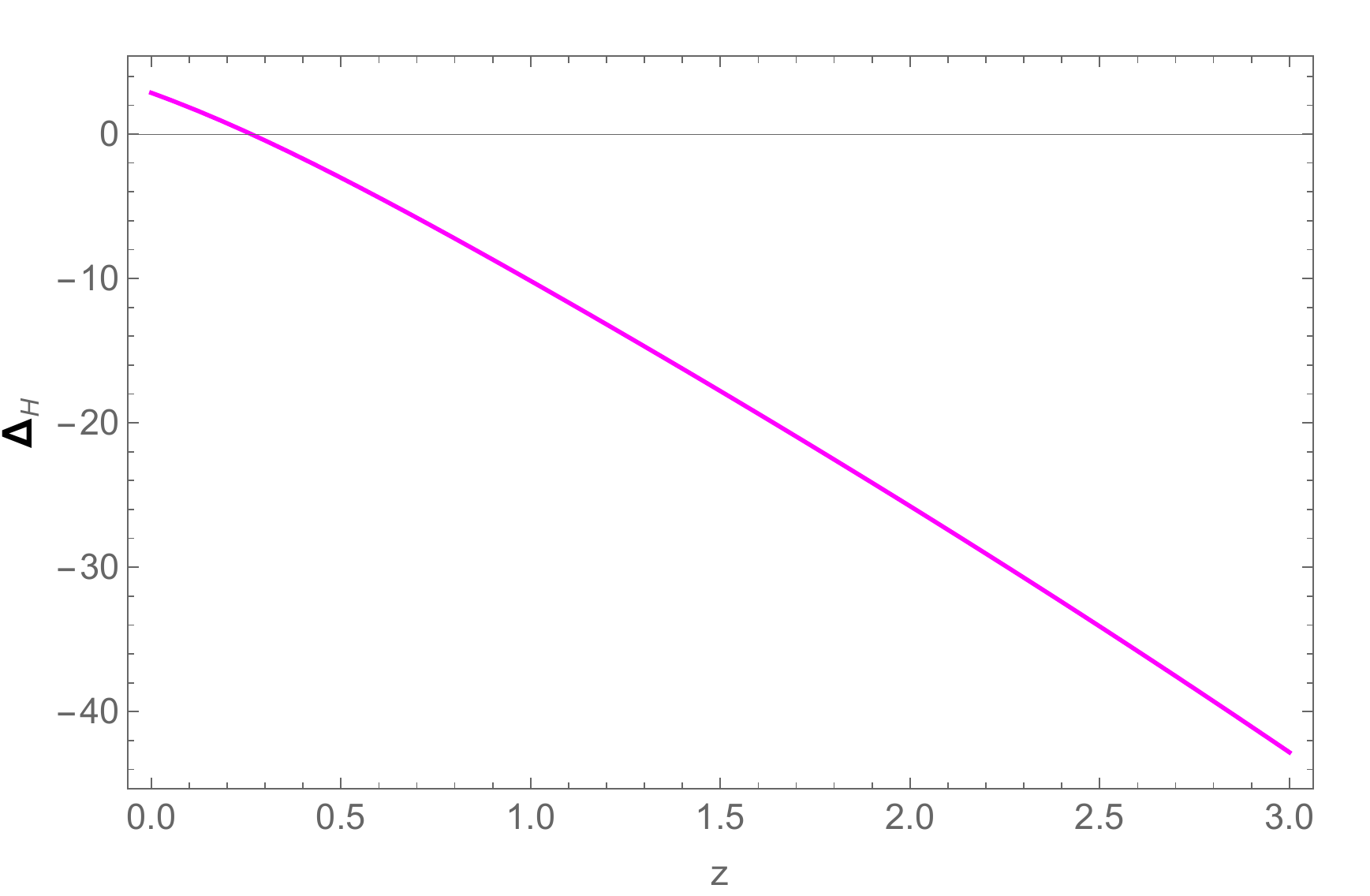}
\includegraphics[scale=0.5]{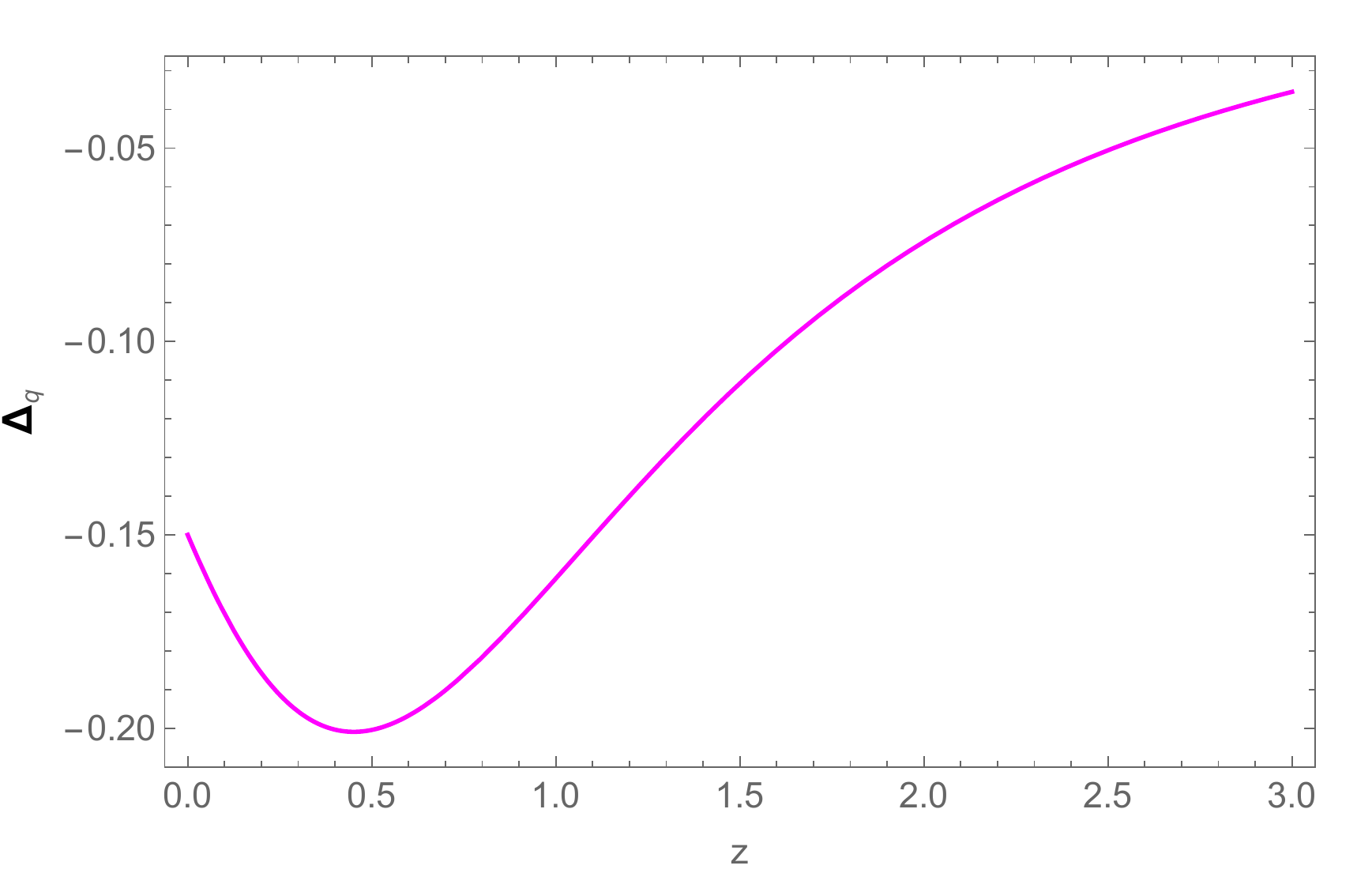}
\caption{Relative Hubble parameter (Top panel) and relative deceleration parameter (Bottom panel) in redshift. The parameter scheme, $\Lambda=e^{3\pi}$.} \label{FIG--2}
\end{figure}
where $\Lambda$ is constant. The Hubble parameter and deceleration parameter of the scale factor can be respectively obtained as, $H=\frac{\dot{a}}{a}= \frac{\sqrt{\Lambda }\coth\left(\frac{\sqrt{3\Lambda}}{2}t\right)}{\sqrt{3}}$ and $ q=-\frac{a\ddot{a}}{{\dot{a}}^2}= \frac{3}{\cosh \left(\sqrt{3\Lambda} t\right)+1}-1$. Also, the slope is obtained as, $\frac{\Lambda }{1-\cosh \left(\sqrt{3\Lambda}t\right)}$. There is always a solution for $\Lambda<0$, that is the cosmos will ultimately collapse. Depending on the relative sizes of the terms, it is possible to find a solution for $\Lambda>0$, but the Universe typically expands indefinitely until its density is high enough to collapse it before the cosmological constant term takes over.
We have given below the graphical behaviour of the Hubble and deceleration parameter in FIG-- \ref{FIG--1}. The Hubble parameter decreases over time and at late time vanishes. The present value of $H$ has been noted as $71.63~kms^{-1}Mpc^{-1}$. The deceleration parameter also decreases over time and is found to be $-0.70$ at present time and at late times it approaches to $-1$. In FIG-- \ref{FIG--2}, we present the relative Hubble parameter $\Delta H=H(z)_{\Lambda CDM}-H(z)$ and the relative deceleration parameter $\Delta q=q(z)_{\Lambda CDM}-q(z)$ as functions of redshift. One may note that, while the magnitude of relative Hubble parameter decreases with time, the magnitude of the relative deceleration parameter initially increases  and after attaining a maximum, it again decreases with the advancement of cosmic time.

\subsection{Case-I: $m=1$}
It is to be noted that, substituting $m=1$ in the above set of equations, one can retrieve the equation of the dynamical parameter for the case $f(Q,T)=\alpha Q+\beta T$. In this case, we have a constant value for the partial derivative of $f$ with respect to the nonmetricity as $F=\alpha$. Consequently, the evolution equation becomes
\begin{equation}
\dot{H}=\frac{4\pi\lambda(1+\omega)}{\alpha}\rho,\label{eq:31}
\end{equation}
where $\lambda=1+\frac{\beta}{4\pi}$. 

For this case with $m=1$, the dynamical parameters of the model may be expressed as,
\newpage
\begin{widetext}
\begin{eqnarray}
\nonumber
p&=&\frac{\alpha}{8\pi(1+2\kappa)}\left[\left(\frac{\sqrt{\Lambda } \coth \left(\frac{\sqrt{3\Lambda}}{2}t\right)}{\sqrt{3}}\right)^{2}\left(3-\left(\frac{3}{1+\cosh \left(\sqrt{3\Lambda} t\right)}\right)(2+\kappa-\kappa\kappa_{1})\right)\right]
\label{Eqn.25}\\
\nonumber
\rho&=&\frac{-\alpha}{8\pi(1+2\kappa)}\left[\left(\frac{\sqrt{\Lambda } \coth \left(\frac{\sqrt{3\Lambda}}{2}t\right)}{\sqrt{3}}\right)^{2}\left(3+\left(\frac{3}{1+\cosh \left(\sqrt{3\Lambda} t\right)}\right)(3\kappa-(2+3\kappa)\kappa_{1})\right)\right]
\label{Eqn.26}\\
\nonumber
\omega&=&-1+\frac{2\left(\frac{3}{1+\cosh \left(\sqrt{3\Lambda} t\right)}\right)(1+2\kappa)(1-\kappa_1)}{3+\left(\frac{3}{1+\cosh \left(\sqrt{3\Lambda} t\right)}\right)\left(3\kappa-\left(2+3\kappa\right)\kappa_1\right)}\label{eq27}
\end{eqnarray}
\end{widetext}

\begin{figure}[H]
\centering
\includegraphics[scale=0.47]{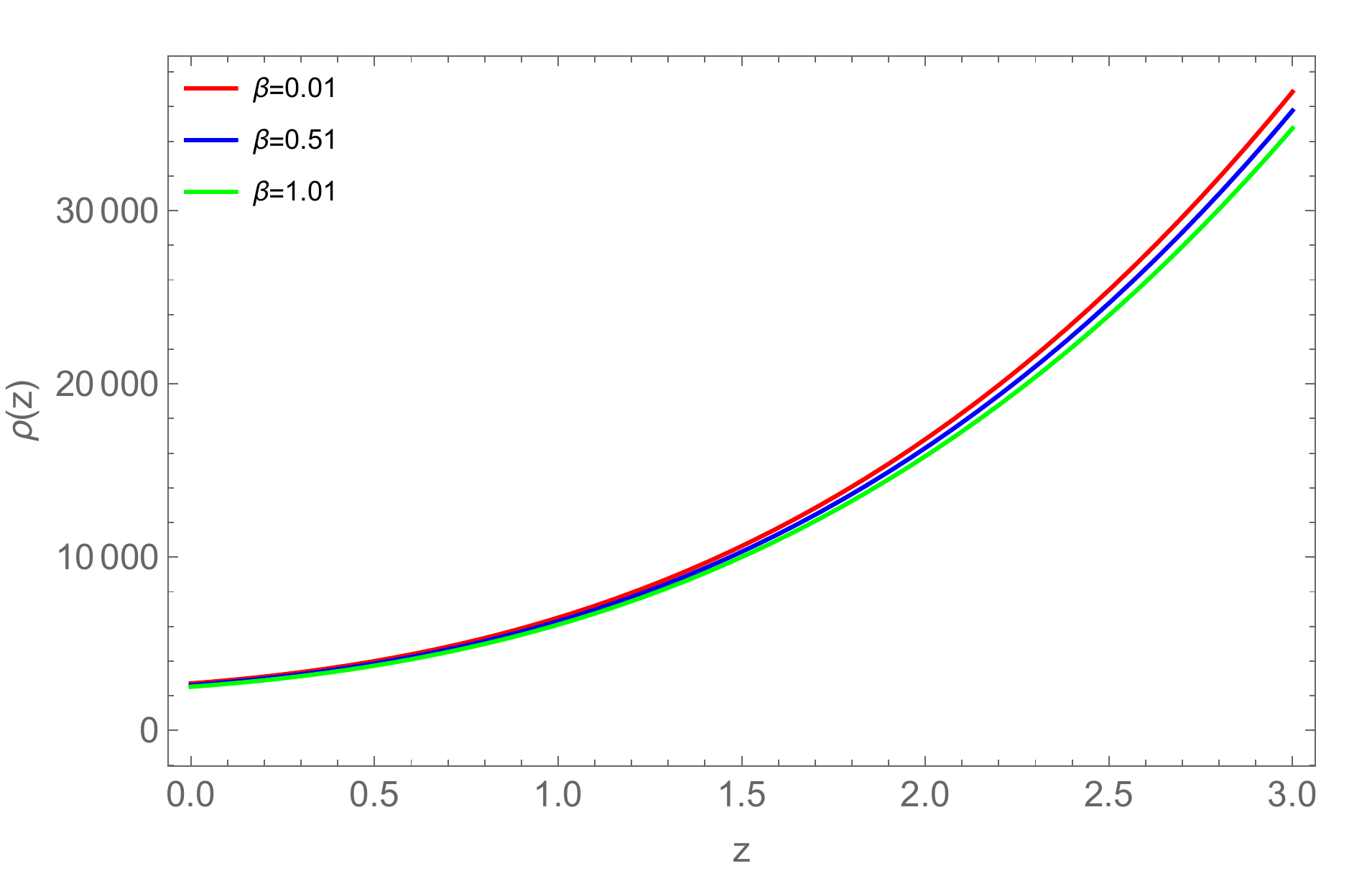}
\includegraphics[scale=0.47]{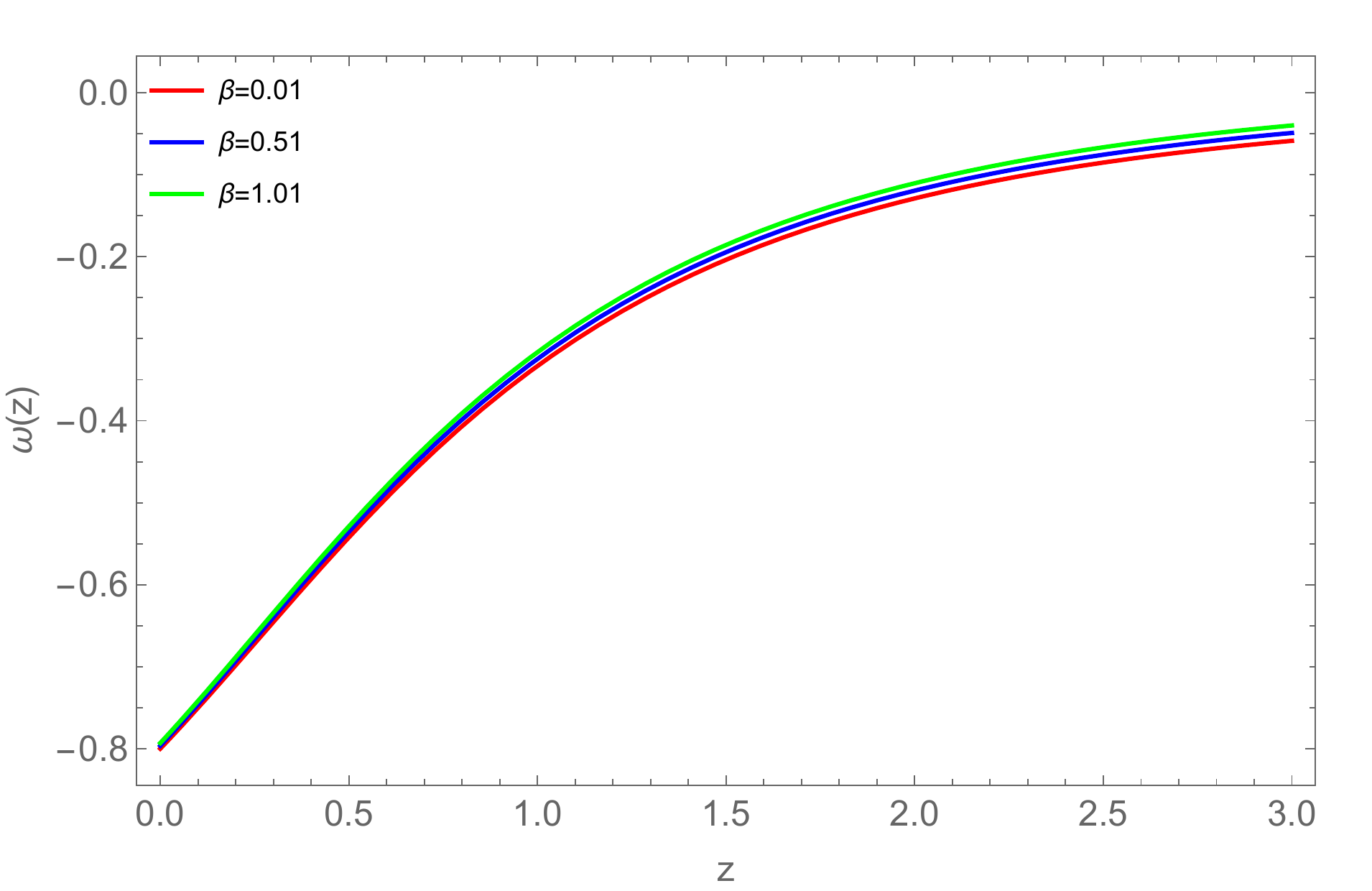}
\caption{Energy density (Top panel) and EoS parameter (Bottom panel) in redshift. The parameter scheme,  $\Lambda=e^{3\pi}$, $\alpha=-4.4$.} \label{FIG--3}
\end{figure}

The graphical behaviour of energy density and EoS parameter has been presented in FIG-\ref{FIG--3}. We have considered here three representative values of the model parameter $\beta=0.01,0.51,1.01$ to assess its impact on the evolution of the curve of energy density and EoS parameter. The other model parameter $\alpha$ kept considered to be a fixed value. The energy density reduces from high positive value to lower one and remain entirely in the positive domain. Lower is the value of the model parameter $\beta$, the evolution starts from higher $\rho$ value and maintain the same behaviour throughout the evolution. Whereas the EoS parameter reduces from higher to lower value entirely in the negative domain. Though a slight variation noticed for the different value of $\beta$
at early epoch, however at late time all merged together and remains at $-1$. This behavior is in agreement with the concordant $\Lambda$CDM model. For the representative values of the parameter $\beta$, at present time ($z=0$), $\omega \in [-0.803,-0.791] $.

\newpage
\begin{widetext}
The effective EoS parameter for $m=1$ becomes
\begin{eqnarray}\label{eq:33}
\omega_{eff}=-1+\frac{4(1+q)}{3-16\pi\rho^{\prime}\left[(1+k)+k\omega\right]}=-1+\frac{4\left(\frac{3}{\cosh \left(\sqrt{3\Lambda} t\right)+1}\right)}{3-16\pi\rho^{\prime}\left[(1+k)+k\omega\right]}
\end{eqnarray}
where $\rho^{\prime}=\frac{\rho}{\alpha H^2}$. Also, the energy conditions can be obtained as,
\begin{eqnarray}
\rho+p&=& \frac{-\alpha}{4\pi}\left(\frac{\sqrt{\Lambda } \coth \left(\frac{\sqrt{3\Lambda}}{2}t\right)}{\sqrt{3}}\right)^{2}\left[ \left(\frac{3}{1+\cosh(\sqrt{3\Lambda}t)}\right)(1-\kappa_{1})\right] \nonumber\\
\rho+3p&=&\frac{-\alpha}{8\pi(1+2\kappa)}\left(\frac{\sqrt{\Lambda } \coth \left(\frac{\sqrt{3\Lambda}}{2}t\right)}{\sqrt{3}}\right)^{2}\left[-3+\left(\frac{3}{1+\cosh(\sqrt{3\Lambda}t)}\right)(3+3\kappa-\kappa_{1}-3\kappa\kappa_{1})\right] \nonumber\\
\rho-p&=&\frac{-\alpha}{8\pi(1+2\kappa)}\left(\frac{\sqrt{\Lambda } \coth \left(\frac{\sqrt{3\Lambda}}{2}t\right)}{\sqrt{3}}\right)^{2}\left[3-2\left(\frac{3}{1+\cosh(\sqrt{3\Lambda}t)}\right)(-1+\kappa-\kappa_{1}-\kappa\kappa_{1})\right] \nonumber
\end{eqnarray}

\begin{figure}[H]
\centering
\includegraphics[scale=0.5]{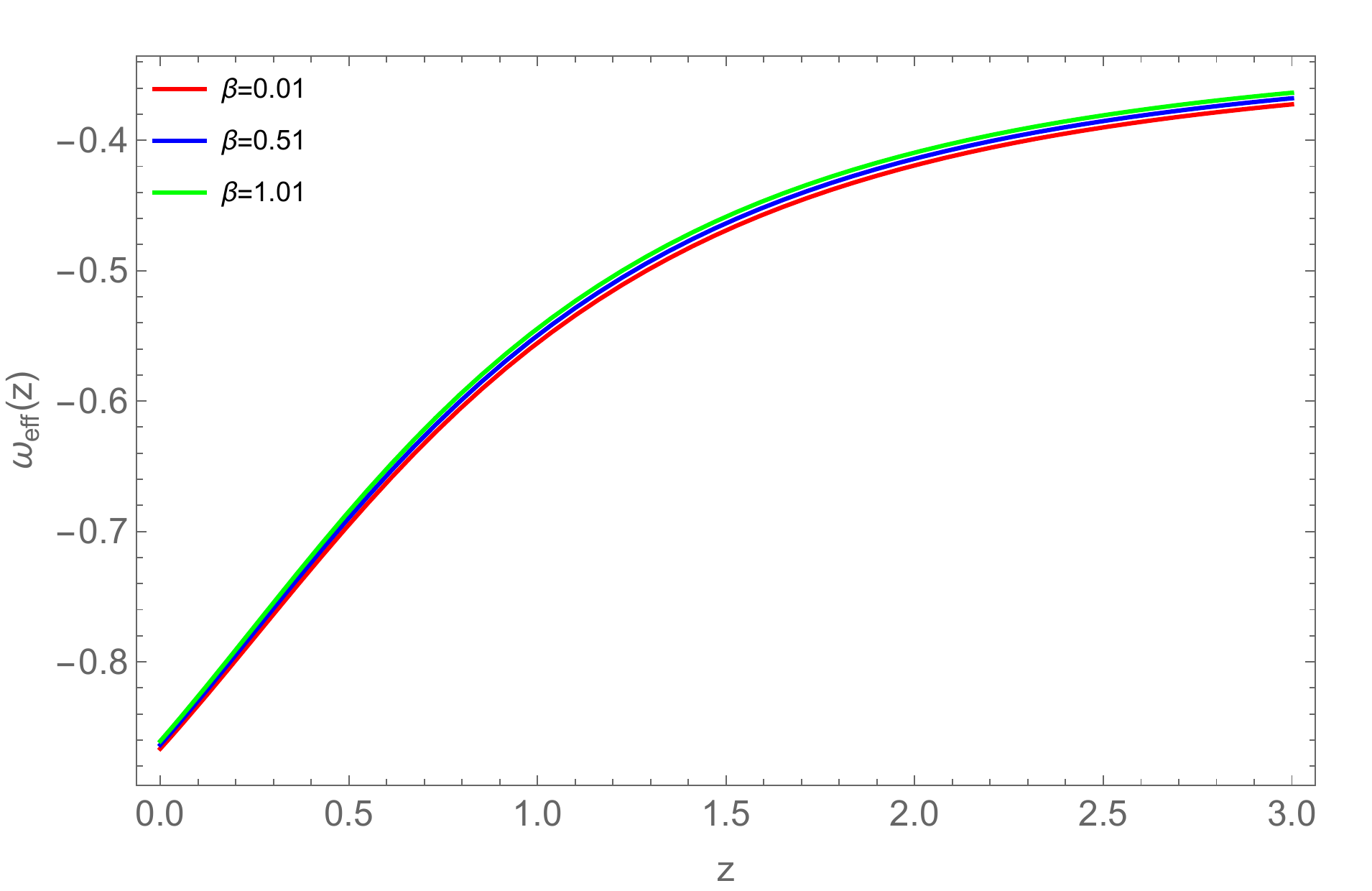}
\includegraphics[scale=0.5]{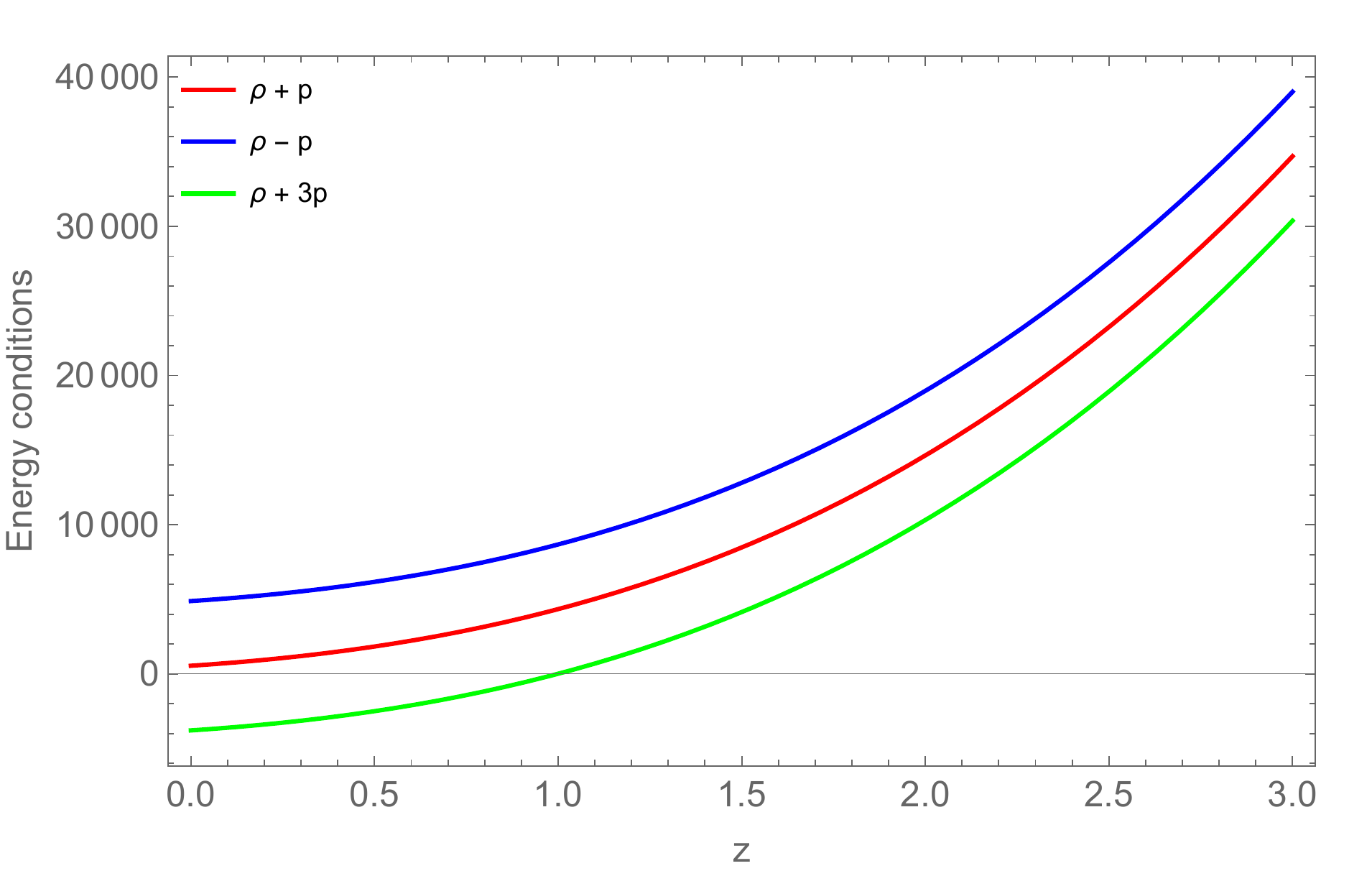}
\caption{Effective EoS parameter (Top panel) and Energy conditions(Bottom panel)  in redshift. The parameter scheme, $\Lambda=e^{3\pi}$, $\alpha=-4.4$. } \label{FIG--4}
\end{figure}
\end{widetext}

FIG-- \ref{FIG--4} represents the effective EoS parameter(left panel) and energy conditions(right panel). The behaviour of effective EoS parameter remains similar to that of EoS parameter except the present value. In this case, the present value is -0.799. All the energy conditions are reducing from higher to lower value. The violation of SEC and positive behaviour of DEC has been noticed, whereas the NEC remains positive and at late times merged with the null line. It confirms the validation of the model.
\subsection{Case-II: $\beta=0$}

With a substitution of $\beta=0$  into the functional $f(Q,T)$ and that in the action, we obtain  the $f(Q)$ gravity with the functional behaving as $f(Q)=\alpha Q^m$. For this specific case, the dynamical parameters of the model reduces to
\begin{widetext}
\begin{eqnarray}
p&=&\frac{1}{16\pi}\left[2^{m}3^{(m-1)}(2m-1)\alpha\left(\frac{\sqrt{\Lambda } \coth \left(\frac{\sqrt{3\Lambda}}{2}t\right)}{\sqrt{3}}\right)^{2m}\left(3-2m\left(\frac{3}{1+\cosh \left(\sqrt{3\Lambda} t\right)}\right)\right)\right]  \label{eq:34}\\
\rho&=&\frac{1}{16\pi}\left[6^m(1-2m)\alpha\left(\frac{\sqrt{\Lambda } \coth \left(\frac{\sqrt{3\Lambda}}{2}t\right)}{\sqrt{3}}\right)^{2m}\right] \label{eq:35}\\
\omega &=&-1+\frac{2}{3}m\left(\frac{3}{1+\cosh \left(\sqrt{3\Lambda} t\right)}\right)\label{eq:36}
\end{eqnarray}
\begin{figure}[H]
\centering
\includegraphics[scale=0.509]{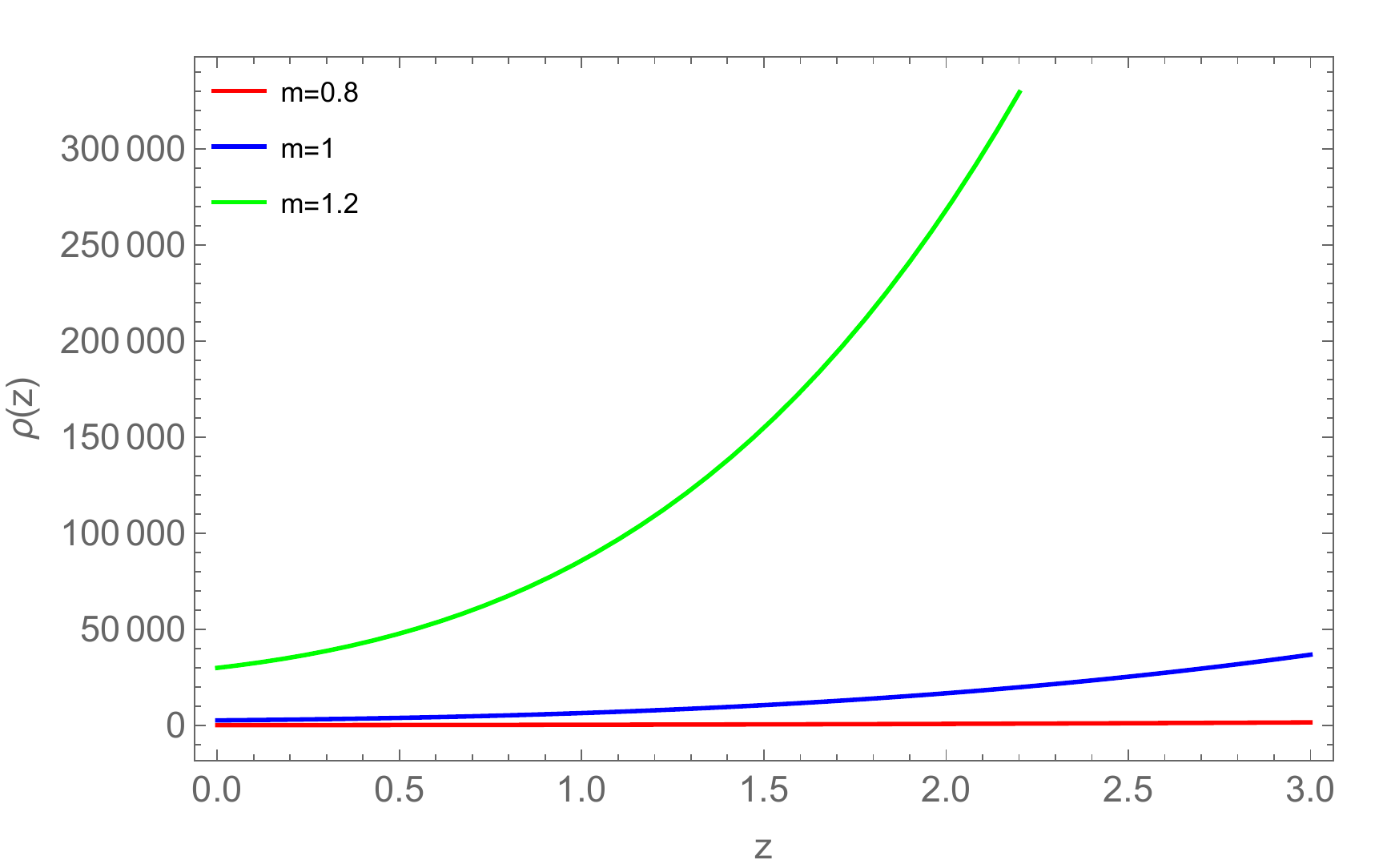}
\includegraphics[scale=0.5]{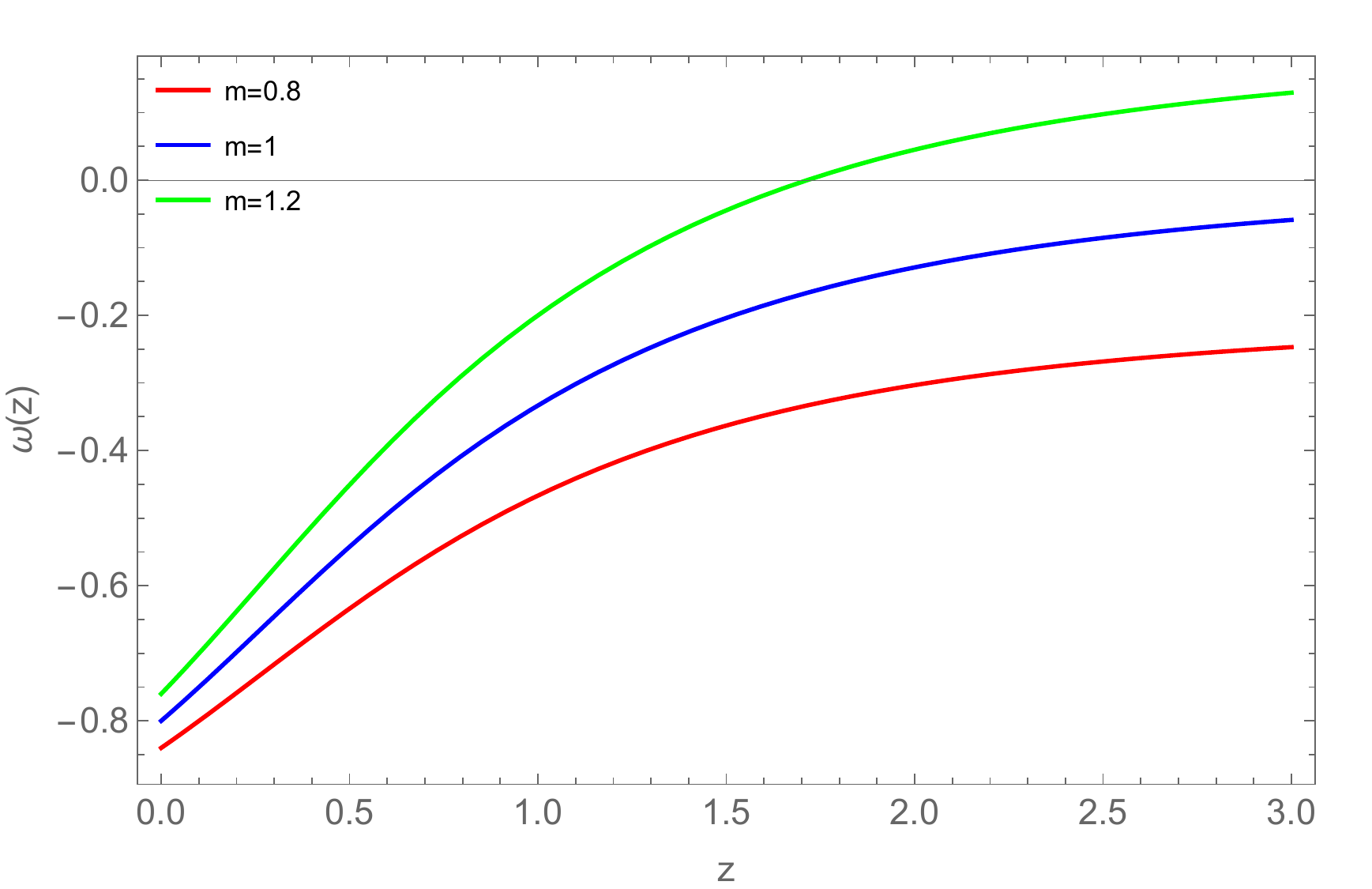}
\caption{Energy density (left panel) and EoS parameter (right panel) in redshift. The parameter scheme,  $\Lambda=e^{3\pi}$, $\alpha=-4.4$.}  \label{FIG--5} 
\end{figure}
\end{widetext}

The evolutionary behaviour of energy density and EoS parameter has been given in FIG-- \ref{FIG--5} for the representative value of the exponent $m=0.6,0.8,1$. For $m=1$, $f(Q)=Q$. The energy density reduces gradually for $m=1$ whereas for $m=0.6$ it remains flat throughout and for $m=0.8$, there is a slight reduction at the start of the evoluion else remains flat throughout. Whereas the EoS parameter reduces from early to late times and merge together at some finite future and approaches to $-1$ at late time. This shows the $\Lambda$CDM behaviour of the Universe. Also higher the value of the exponent $\beta$, the evolution starts from higher value of EoS parameter. At present, the EoS value has been recorded in the range $[-0.878,-0.802]$. Now, the the effective EoS parameter becomes
\begin{widetext}
\begin{eqnarray}\label{eq:37}
\omega_{eff}=-1+\frac{2^{m+2}3^{m-1}\alpha mH^{2m}(1+q)}{\alpha 2^{m}3^{m}H^{2m}-16\pi\rho}=-1+\frac{2^{m+1}3^{m-1}\alpha m\left(\frac{\sqrt{\Lambda } \coth \left(\frac{\sqrt{3\Lambda}}{2}t\right)}{\sqrt{3}}\right)^{2m}\left(\frac{3}{1+\cosh \left(\sqrt{3\Lambda} t\right)}\right)}{3^{m} 2^{m-1}\alpha \left(\frac{\sqrt{\Lambda } \coth \left(\frac{\sqrt{3\Lambda}}{2}t\right)}{\sqrt{3}}\right)^{2m}-8\pi\rho}
\end{eqnarray}

\begin{figure}[!htp]
\centering
\includegraphics[scale=0.5]{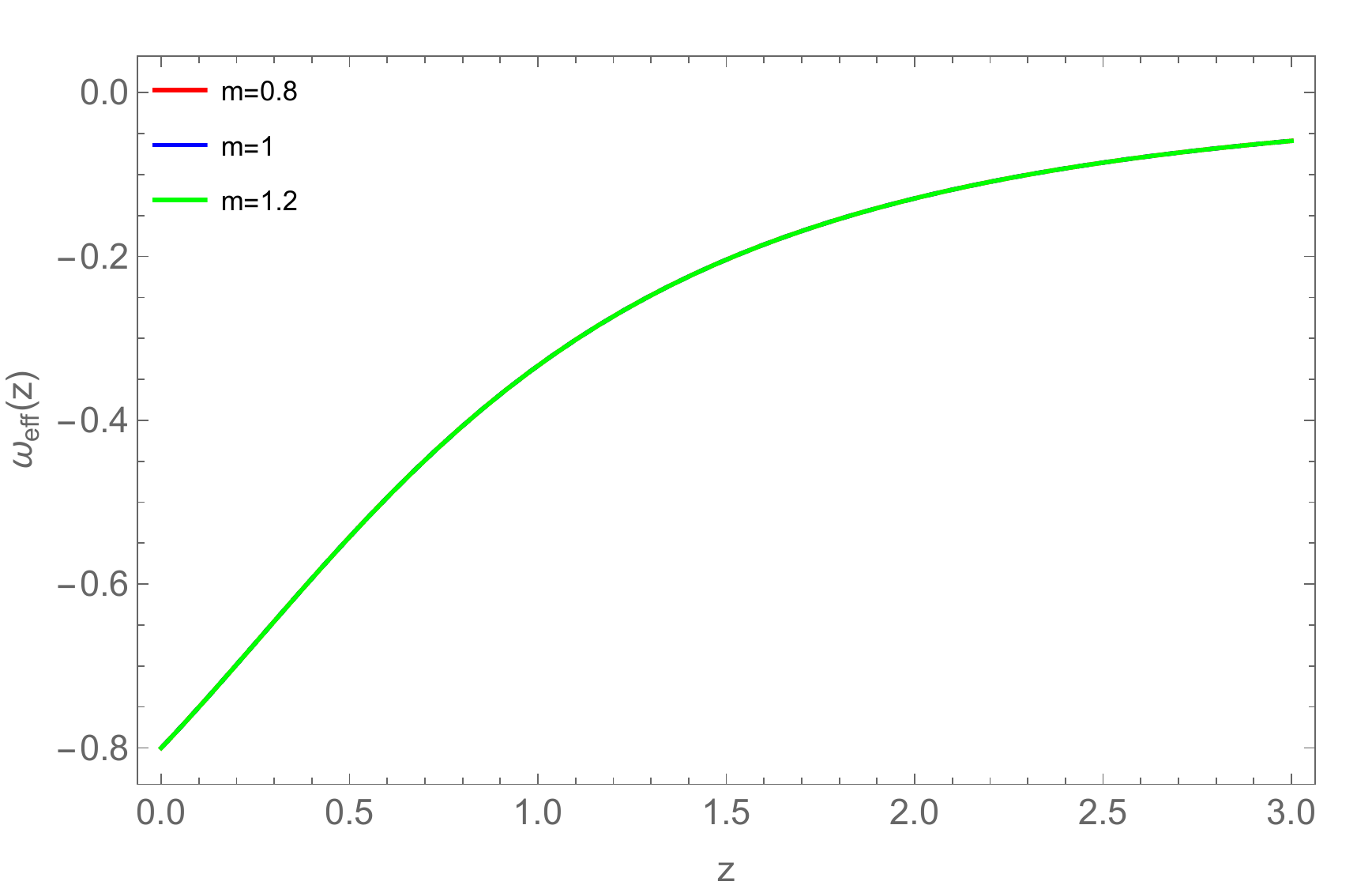}
\includegraphics[scale=0.5]{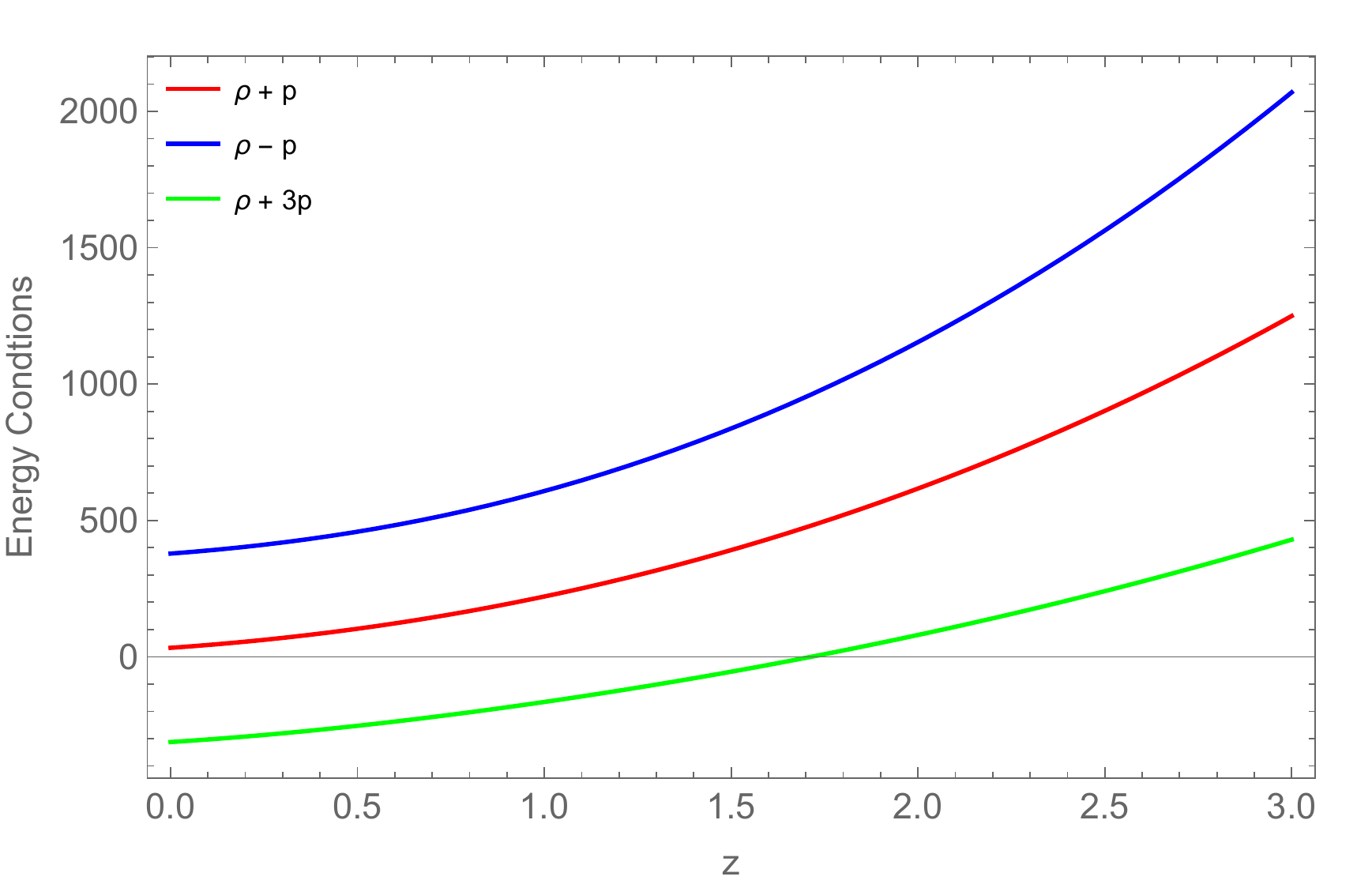}
\caption{Effective EoS parameter (left panel) and Energy conditions (right panel) in redshift. The parameter scheme,  $\Lambda=e^{3\pi}$, $\alpha=-4.4$. } \label{FIG--6}  
\end{figure}
and the energy conditions for Case-II becomes,
\begin{eqnarray}\label{eq:38}
\rho+p &=&\frac{1}{16\pi}\left[2^{m+1}3^{m-1}\alpha m(1-2m)\left(\frac{\sqrt{\Lambda } \coth \left(\frac{\sqrt{3\Lambda}}{2}t\right)}{\sqrt{3}}\right)^{2m}\left(\frac{3}{1+\cosh \left(\sqrt{3\Lambda} t\right)}\right)\right], \nonumber \\
\rho+3p &=&\frac{1}{16\pi}\left[2^{m+1}3^{m}\alpha(1-2m)\left(\frac{\sqrt{\Lambda } \coth \left(\frac{\sqrt{3\Lambda}}{2}t\right)}{\sqrt{3}}\right)^{2m}\left(-1+m\left(\frac{3}{1+\cosh \left(\sqrt{3\Lambda} t\right)}\right)\right)\right],\nonumber \\
\rho-p &=&\frac{1}{16\pi}\left[2^{m+1}3^{m-1}\alpha(1-2m)\left(\frac{\sqrt{\Lambda } \coth \left(\frac{\sqrt{3\Lambda}}{2}t\right)}{\sqrt{3}}\right)^{2m}\left(3-m\left(\frac{3}{1+\cosh \left(\sqrt{3\Lambda} t\right)}\right)\right)\right] \label{EC2}
\end{eqnarray}
\end{widetext}
The behaviour of effective EoS shows the similar behavior to that of EoS, in fact the present value of $\omega_{eff}=-0.802$ is the same as that of EoS for $m=1$. It indicates that the dark energy phase dominates the evolution and the matter part becomes suppressed. For different values of $m$, no change has been noticed and all are lying on the same curve. The energy conditions for Case-II also show the similar behaviour that of Case-I. The DEC satisfies, NEC initially satisfies and over the time merged with the null line and at infinite future shows violation. The SEC violates entirely and this behaviour has been inevitable in modified theories of gravity.

\section{Dynamical System Analysis} \label{Sec.V}
To analyze the stability of the system, we shall perform the dynamical system analysis, which requires recasting the cosmological equations in terms of dynamical systems. Here, we shall present the cosmological dynamical system in $f(Q, T)$ gravitational theory. We consider $f(Q, T) = Q + \Phi(Q,T)$, then Eqs. \eqref{eq:10} and \eqref{eq:11} become,
\begin{widetext}
\begin{eqnarray}
    3H^{2} &=& -8\pi\rho+\frac{\Phi}{2}-Q\Phi_{Q}-\frac{2\Phi_{T}}{8\pi+\Phi_{T}}\left[(1+\Phi_{Q}+2Q\Phi_{QQ})\dot{H}+\Phi_{QT}H\dot{T} \right]~,\label{eq:39}\\
    2\dot{H} &=& \frac{8\pi p+\frac{\Phi}{2}-Q\Phi_{Q}-2\Phi_{QT}H\dot{T}-3H^{2}}{1+\Phi_{Q}+2Q\Phi_{QQ}}~.\label{eq:40}
\end{eqnarray}

We consider the Universe is filled with dust and radiation fluid and therefore,
\begin{equation*}
    \rho=\rho_{m}+\rho_{r}~,~~~~~~~~~~ p=p_{m}+p_{r}~,~~~~~~~~~~ T=-\rho+3p~.
\end{equation*}
where $\rho_{r}$ and $\rho_{m}$ are the energy densities of radiation and matter respectively. The geometrical dark energy [$\Phi(Q, T)$] density and pressure can be expressed as,

\begin{eqnarray}
    \rho_{de} &=& \frac{\Phi}{2}-Q\Phi_{Q}-\frac{2\Phi_{T}}{8\pi+\Phi_{T}}\left[(1+\Phi_{Q}+2Q\Phi_{QQ})\dot{H}+\Phi_{QT}H\dot{T} \right]~,\label{eq:41}\\
    p_{de} &=& -\frac{\Phi}{2}+Q\Phi_{Q}+2\Phi_{QT}H\dot{T}+2\dot{H}(\Phi_{Q}+2Q\Phi_{QQ})~.\label{eq:42}
 \end{eqnarray}
 \end{widetext}
Now, we may introduce the density parameter pertaining to pressureless matter, radiation and dark energy respectively as,
\begin{equation}\label{eq:43}
\Omega_{m} = \frac{\rho_{m}}{3H^{2}},~~~~~~~~~~~~~\Omega_{r} = \frac{\rho_{r}}{3H^{2}}~.,~~~~~~~~~~~~~\Omega_{de} = \frac{\rho_{de}}{3H^{2}}~.
\end{equation}
To analyse the dynamics of the cosmological models, we use the following dimensionless variables $x$, $y$, $z$, $u$ and $v$. Let $\kappa^{2} = -8\pi$ then Eqns. \eqref{eq:39}-\eqref{eq:40} and conservation equation are transformed into an autonomous system of first-order differential equations
\begin{eqnarray}\label{eq:44}
    x=\frac{\kappa^{2}\rho_{m}}{3H^{2}}~,\hspace{0.5cm} y=\frac{\kappa^{2}\rho_{r}}{3H^{2}}~,\hspace{0.5cm} z=\frac{\Phi}{6H^{2}}~\nonumber\\
    u=-2\Phi_{Q}~,\hspace{0.5cm}v=\frac{\Phi_{T}}{\Phi_{T}-\kappa^{2}}.
\end{eqnarray}
Therefore, Eqns. \eqref{eq:39}-\eqref{eq:40} and conservation equation can be transformed  into following dynamical system,
\begin{eqnarray}
    x' &=& -x\left(3 + 2\frac{\dot{H}}{H^{2}}\right)~,\label{eq:45}\\
    y' &=& -2y\left( 2+\frac{\dot{H}}{H^{2}}\right)~'\label{eq:46}\\
    z' &=& -\left(2z\frac{\dot{H}}{H^{2}}+6u\frac{\dot{H}}{H^{2}}+\frac{3xv\kappa^{4}}{2(v-1)} \right)~,\label{eq:47}\\
    u' &=& -24\dot{H}\Phi_{QQ}-6\rho_{m}\Phi_{QT}~,\label{eq:48}\\
    v' &=& \left(\frac{v(1-v)}{\Phi_{T}}\right)(12\dot{H}\Phi_{TQ}-3\rho_{m}\Phi_{TT})~.\label{eq:49}
\end{eqnarray}
where prime denotes a derivative with respect to $ln~a$ and
\begin{eqnarray*}
    \frac{\dot{H}}{H^{2}} &=& \frac{3z+3u-y-3+3x\kappa^{2}Q\Phi_{QT}}{2(2Q\Phi_{QQ}+\Phi_{Q}+1)}~,\\
    \omega_{de} &=& -\frac{-\frac{(2 m (u-2)-u+4) (3 u-y+3 z-3)}{2 m (u-1)-u+2}+u+z}{u (v+1)+v \left(-\frac{y}{3}+z-1\right)+z}~,\\
    \Omega_{m} &=& x~, \hspace{0.5cm}\Omega_{r}=y~.
\end{eqnarray*}
To analyse the models, we have to consider some forms  of $\Phi(Q,T)$.\\
{\bf Form I:} We consider, $\Phi(Q, T)= \alpha Q^{m}+\beta T -Q$, then
\begin{equation}\label{eq:50}
    \frac{\dot{H}}{H^{2}} = \frac{3z+3u-y-3}{u(1-2m)+2(m-1)}
\end{equation}
Subsequently, Eqns. \eqref{eq:45}-\eqref{eq:49}  reduce to, 
\begin{eqnarray*}
    x' &=& \frac{x}{u(1-2m)+2(m-1)}(2y-6z-9u+6mu-6m+12)~,\label{eq:51a}\\
    y' &=& \frac{2y}{u(1-2m)+2(m-1)}(y-3z-5u+4mu-4m+7)~,\label{eq:52a}\\
    z' &=& -\left( \frac{3z+3u-y-3}{u(1-2m)+2(m-1)}(2z+6u)+\frac{3\beta\kappa^{2}}{2}x\right)~,\label{eq:53a}\\
    u' &=& (m-1)(u-2)\left(\frac{3z+3u-y-3}{u(1-2m)+2(m-1)}\right)~,\label{eq:54a}\\
    v' &=& 0~.
\end{eqnarray*}

\begin{figure}[H]
\centering
\includegraphics[scale=0.5]{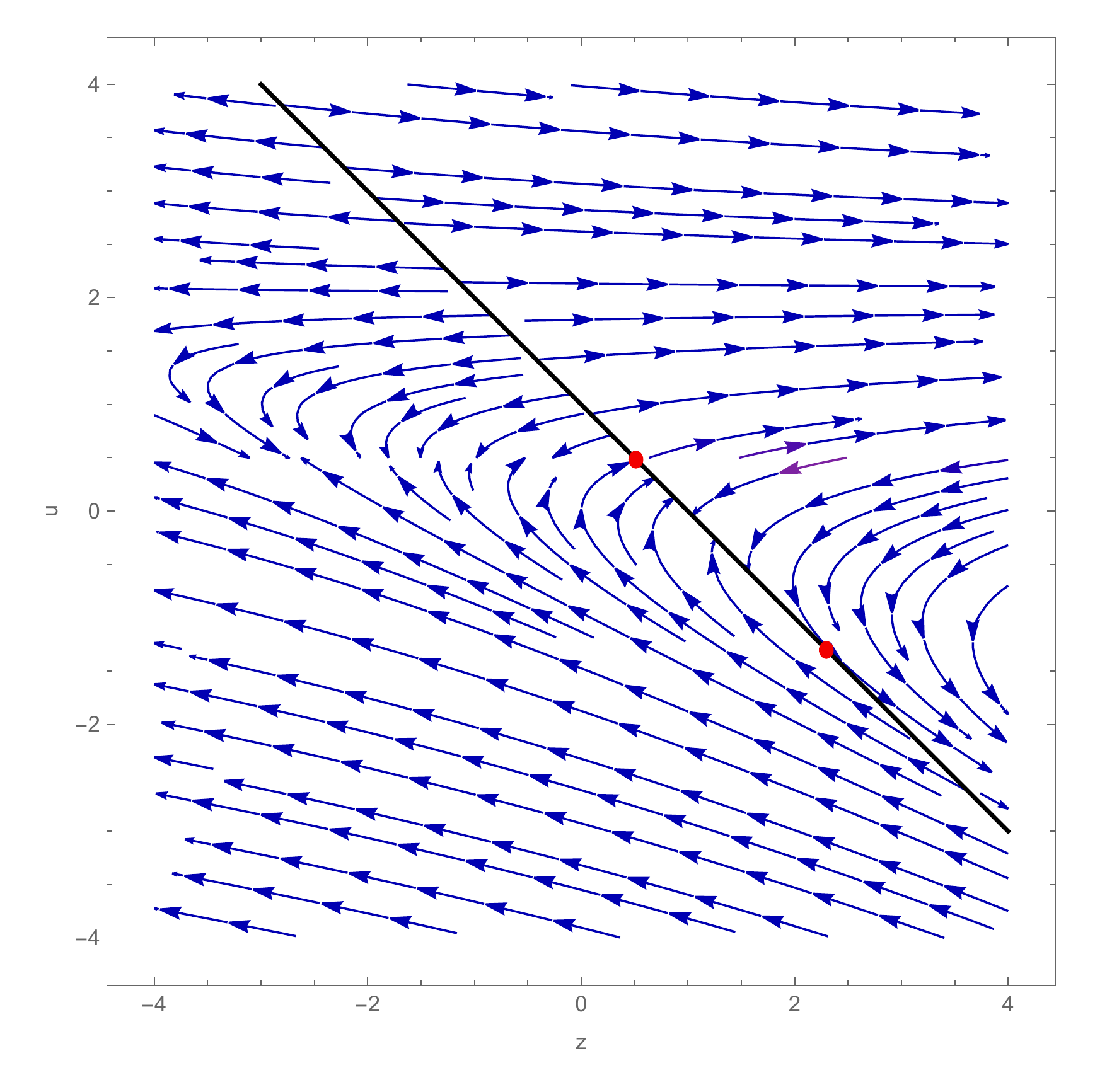}
\caption{Phase portrait for the dynamical system for the form $\Phi(Q,T) = \alpha Q^{m}+\beta T -Q$,~~( $m=1.5$).} \label{FIG--7}
\end{figure}

\begin{widetext}
The term $v'=0$ further reduce the above system as,
\begin{eqnarray}
    x' &=& \frac{x}{u(1-2m)+2(m-1)}(2y-6z-9u+6mu-6m+12)~,\label{eq:51}\\
    y' &=& \frac{2y}{u(1-2m)+2(m-1)}(y-3z-5u+4mu-4m+7)~,\label{eq:52}\\
    z' &=& -\left( \frac{3z+3u-y-3}{u(1-2m)+2(m-1)}(2z+6u)+\frac{3\beta\kappa^{2}}{2}x\right)~,\label{eq:53}\\
    u' &=& (m-1)(u-2)\left(\frac{3z+3u-y-3}{u(1-2m)+2(m-1)}\right)~.\label{eq:54}
\end{eqnarray}
\end{widetext}
In order to derive the dynamical features of the autonomous system, the coupled equation $x'=0$, $y'=0$, $z'=0$ and $u'=0$ are to be solved. TABLE- \ref{table:I} shows the corresponding critical points and their descriptions for the above system. In TABLE- \ref{table:II}, stability conditions and the corresponding cosmological parameters are given. 
\begin{widetext}

\begin{table}[H]
\begin{center}
\renewcommand{\arraystretch}{2.5}
\caption{Critical points for the dynamical system.}\label{table:I}
\begin{tabular}{|c|c|c|c|c|c|}
\hline\hline
 Critical Point &$x$ & $y$ & $z$ & $u$ &  Exists For\\
 \hline\hline
   $A_1$ & $0$ & $0$ & $\lambda$ & $1-\lambda$ &  $2m\lambda-\lambda-1\neq 0$  \\\hline
   $B_1$ & $0$ & $-15-4m$ & $-6$ & $2$ &  $m\neq 0$  \\\hline
   $C_1$ & $\gamma$ & $0$ & $-6$ & $2$ &  $m=-5~~\&~~\kappa\neq 0~~\&~~\beta==0$  \\\hline
   $D_1$ & $\frac{2(5+m)}{\beta\kappa^{4}}$ & $0$ & $m-1$ & $2$ & $m\neq0~~\&~~\beta\neq 0$ \\\hline
   $E_1$ & $0$ & $0$ & $\lambda_{1}$ & $1-\lambda_{1}$ & $\lambda_{1}\neq 1$~~\&~~$m=1$\\\hline
   $F_1$ & $0$ & $\xi_{1}$ & $\frac{3(3+\xi_{1})}{8}$ & $\frac{-3-\xi_{1}}{8}$ & $\xi_{1}\neq -3$~~\&~~$m=1$ \\
   \hline
\end{tabular}
\end{center}
\end{table}

\begin{table}[H]
\begin{center}
\renewcommand{\arraystretch}{2.0}
\caption{Stability conditions, EoS parameter and deceleration parameter}\label{table:II}
\begin{tabular}{|c|c|c|c|c|}
\hline\hline
   Critical Point & Stability Conditions  & $q$ & $\omega_{eff}$ & $\omega_{de}$ \\
   \hline\hline
   $A_1$ & \begin{tabular}{@{}c@{}} Stable for \\ $m<\frac{1}{2} \left(-\sqrt{14}-2\right)\land \left(\alpha <\frac{-m-2}{m-3}\lor \alpha >\frac{1}{2m-1}\right)$,\\$\frac{1}{2} \left(-\sqrt{14}-2\right)<m<\frac{1}{2}\land \left(\alpha <\frac{1}{2m-1}\lor \alpha >\frac{-m-2}{m-3}\right)$,\\ $\left(m=\frac{1}{2}\land \alpha >1\right)\lor \left(\frac{1}{2}<m<\frac{1}{2} \left(\sqrt{14}-2\right)\land \frac{-m-2}{m-3}<\alpha <\frac{1}{2 m-1}\right)$,\\ $\frac{1}{2} \left(\sqrt{14}-2\right)<m<3\land \frac{1}{2m-1}<\alpha <\frac{-m-2}{m-3}$,\\ $\left(m=3\land \alpha >\frac{1}{5}\right)\lor \left(m>3\land \left(\alpha <\frac{-m-2}{m-3}\lor \alpha >\frac{1}{2m-1}\right)\right)$  \end{tabular} & $-1$ & $-1$ & $-1$ \\\hline
   $B_1$ & Unstable & $1$ & $\frac{1}{3}$ & $\frac{6}{m-3}$ \\\hline
   $C_1$ & $Unstable$ & $\frac{1}{2}$ & $0$ & $\frac{-7}{9}$ \\\hline
   $D_1$ & Unstable & $\frac{1}{2}$ & $0$ & $\frac{2-m}{1+2m}$ \\\hline
   $E_1$ & \begin{tabular}{@{}c@{}} Stable for \\
   $1<\lambda_{1} <\frac{3}{2}$ \end{tabular}& $-1$ & $-1$ & $-1$ \\\hline
   $F_1$ & Unstable & $1$ & $\frac{1}{3}$ & $-3$ \\
   \hline
\end{tabular}
\end{center}
\end{table}
\end{widetext}
An important tool in the study of dynamical systems is the phase portrait, which plots typical trajectory plots. The phase portrait can be used to determine the stability of the models. FIG---\ref{FIG--7} shows the phase space portrait diagram for the dynamical system Eqs. \eqref{eq:51}-\eqref{eq:54} in $2-D$, $u$ vs $z$ plane. It can be seen that for $m=1.5$, the line $(z,1-z)$ is stable in the interval $\frac{1}{2}<z<\frac{7}{3}$. The description of each critical points are given below:  
\begin{itemize}
 \item \textbf{Critical Point $A_{1}$ :} At this point, $\Omega_{de}=1$, $\Omega_{m}=0$ and $\Omega_{r}=0$, i.e the Universe shows dark energy dominated phase. The accelerated dark energy dominated Universe is confirmed by the corresponding values of the EoS parameter ($\omega_{eff}=-1$) and deceleration parameter $q=-1$. The eigenvalues for the Jacobian matrix using the critical points are negative real part and zero. Further, there is only one vanishing eigenvalue and therefore the dimension of the set of eigenvalues equals its number. The critical point associated with it cannot be a global attractor \cite{Aulbach84, Coley99}.   At this critical point, the stable node is resulted. 
 \begin{eqnarray*}
 \left\{0,~-\frac{6 (-3\lambda+\lambda m+m+2)}{-\lambda+2\lambda m-1},~-4,~-3\right\}.
\end{eqnarray*}

\item \textbf{Critical Point $B_{1}$ :} At this point that the deceleration parameter and EoS parameter are respectively $q=1$ and $\omega_{eff}=\frac{1}{3}$, which indicates the decelerating phase of the Universe. The density parameters are $\Omega_{m}=0$, $\Omega_{r}=-15-4m$ and $\Omega_{de}=0$. For $m=-4$, the Universe shows radiation dominated phase. This critical point is an unstable saddle because it contains both negative and positive eigenvalues of the Jacobian matrix.
 \begin{equation*}
 \left\{1,~4,~-4(n-1),~\frac{4 n+15}{n}\right\}.
 \end{equation*}

 \item \textbf{Critical Point $C_{1}$ :} This point exists for $\gamma\neq 0$ and the corresponding deceleration parameter $q = \frac{1}{2}$ and EoS parameter $\omega_{eff}=0$. This behaviour of the critical point leads to the decelerating phase of the Universe. Also, density parameters are $\Omega_{m}=\gamma$, $\Omega_{r}=0$ and $\Omega_{de}=1-\gamma$. If we consider $\gamma=1$, then the Universe shows matter dominated phase. The eigenvalues of the Jacobian matrix for this critical point are given below. There is a positive and negative signature of the eigenvalues, showing unstable saddle behavior.
    \begin{equation*}
        \left\{18,-1,\frac{3}{10} \left(5- \sqrt{10\beta\kappa^4\gamma +25}\right),\frac{3}{10}\left(5+\sqrt{10\beta \kappa^4\gamma +25}\right)\right\}.
    \end{equation*}

    \item \textbf{Critical Point $D_{1}$ :} This point exists for $m\neq 0$ and $\beta\neq 0$. The corresponding deceleration parameter is $q=\frac{1}{2}$ and EoS parameter is $\omega_{eff}=0$. The density parameters are $\Omega_{m}=\frac{2(5+m)}{\beta\kappa^{4}}$, $\Omega_{r}=0$ and $\Omega_{de}=0$. For $m=-4$ and $\frac{2}{\beta\kappa^{4}}=1$, the Universe shows the matter dominated phase. This behaviour of the critical point leads to the decelerating phase of the Universe and the eigenvalues of Jacobian matrix for these critical points are positive and negative signature, shows unstable saddle behavior.
    \begin{equation*}
        \left\{-1,~3,~-3(m-1),~\frac{3(m+5)}{m}\right\}.
    \end{equation*}

 \item \textbf{Critical Point $E_{1}$ :} This point exists for $m=1$ and $1 < \lambda_{1} < \frac{3}{2}$. The deceleration parameter and EoS parameter are $q=-1$ and $\omega_{eff}=-1$ respectively. The values of density parameters are $\Omega_{m}=0$, $\Omega_{r}=0$ and $\Omega_{de}=1$ and the Universe showing the dark energy dominated phase. The eigenvalues of Jacobian matrix for critical points are negative real part and zero gives stable node and accelerating phase of the Universe. The eigenvalues are given below.
 \begin{equation*}
 \left\{0,~-4,~-3,~\frac{6(2\lambda_{1}-3)}{\lambda_{1}-1}\right\}.
 \end{equation*}

 \item \textbf{Critical Point $F_{1}$ :} The points exists for $m=1$ and $\xi_{1}\neq -3$. The deceleration parameter is $q=1$ and EoS parameter is $\omega_{eff}=\frac{1}{3}$. The density parameters are $\Omega_{m}=0$, $\Omega_{r}=\xi_{1}$ and $\Omega_{de}=0$ and is we consider $\xi_{1}=1$  shows the radiation dominated phase of the Universe. This behaviour of the critical point leads to the decelerating phase of the Universe and the eigenvalues of Jacobian matrix for these critical points are zero and positive signature, shows unstable node behavior.
 \begin{equation*}
\left\{0,~\frac{16\xi_{1}}{\xi_{1}+3},~1,~4\right\}.
\end{equation*}
\end{itemize}

As we can see that the stability of the model $\Phi(Q,T)=\alpha Q^{m}+\beta T -Q$ is depends on the value of $n$. If the $n$ value is not equal to $1$, then the critical point $A_{1}$ shows that there are different intervals for $n$ where we get stable behavior of the model in different planes. And if $n=1$, then we have a plane of stability $(0,~0,~\lambda_{1},~1-\lambda_{1})$ shown at critical point $E_{1}$ with some condition on $\lambda_{1}$. Moreover there are some unstable critical points $B_{1}$, $C_{1}$, $D_{1}$ and $F_{1}$, among them $B_{1}$ and $F_{1}$ gives radiation dominated phase of the Universe, while $C_{1}$ and $D_{1}$ gives matter dominated phase of the Universe.\\

\begin{figure}[H]
    \centering
    \includegraphics[scale=0.5]{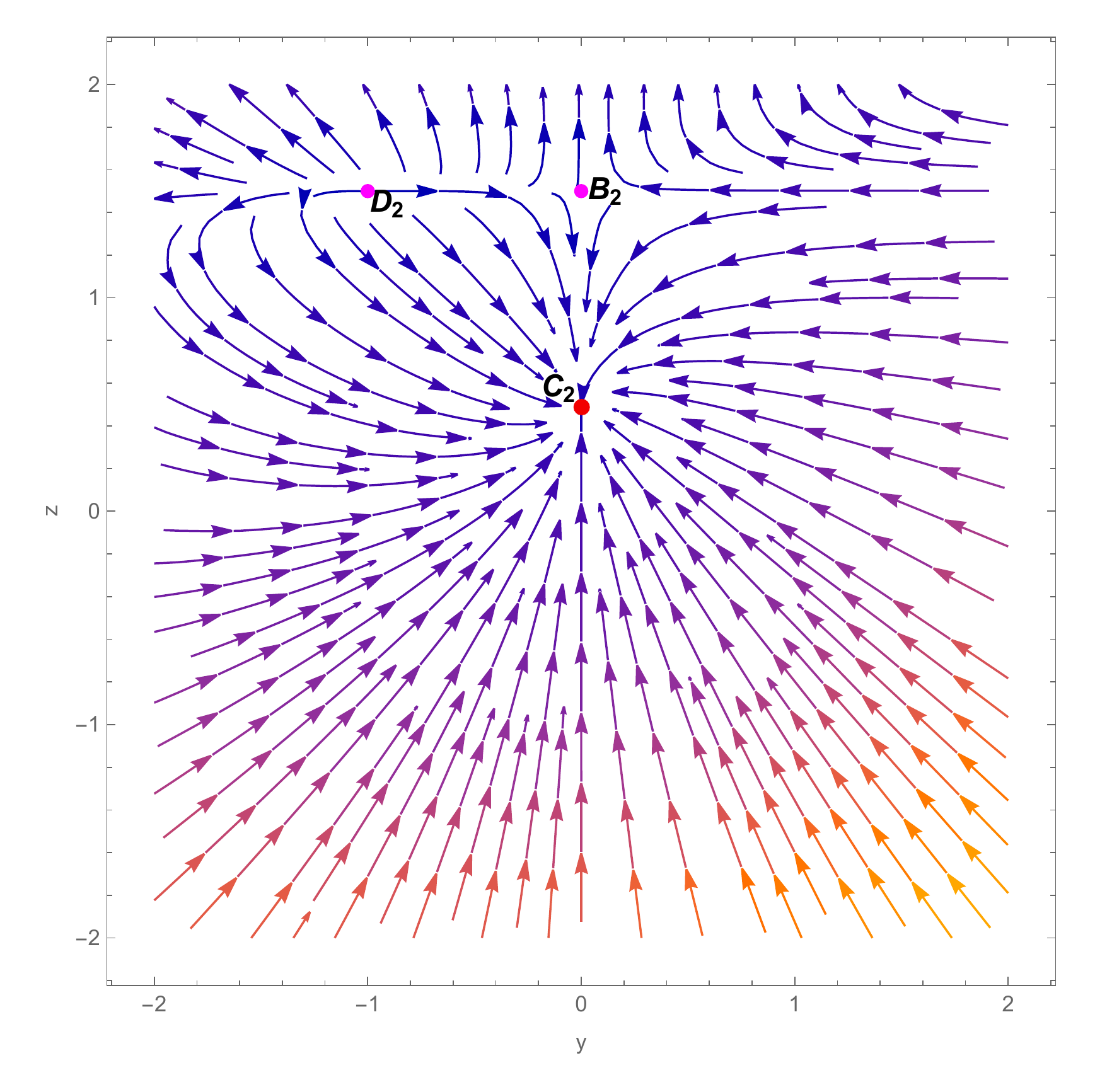}
    \caption{Phase portrait for the dynamical system for the form, $\Phi(Q,T) = \alpha_{1}Q+\beta_{1}T-Q$ with $\alpha_{1}=-1$.}
    \label{FIG--8}
\end{figure}

{\bf Form II:} We consider the form as, $\Phi(Q, T)= \alpha_{1}Q+\beta_{1}T^{2}-Q$, then 
\begin{eqnarray*}\label{eq:55}
\frac{\dot{H}}{H^{2}} &= \frac{3z+3u-y-3}{2\alpha_{1}}~,
\end{eqnarray*}
\begin{eqnarray}
\omega_{de} &= \frac{-9 \alpha_{1}^2+15 \alpha_{1}+2 (\alpha_{1}-1)y+6z-6}{2 \alpha_{1}(v(6\alpha_{1}+y-3 z-3)-3 (-2\alpha_{1}+z+2))}~.
\end{eqnarray}
Now, Eqns. \eqref{eq:45}-\eqref{eq:49} become,
\begin{eqnarray*}
x' &=& \frac{x(y-3z-3u-3\alpha_{1}+3)}{\alpha_{1}}~,\label{eq:56a}\\
y' &=& \frac{y(y-3z-3u-4\alpha_{1}+3)}{\alpha_{1}}~,\label{eq:57a}\\
z' &=& -\left(\frac{3z+3u-y-3}{2\alpha_{1}}(2z+6u)+\frac{3\kappa^{4}}{2}\frac{xv}{v-1}\right)~,\label{eq:58a}\\
u' &=& 0~,\label{eq:59a}\\
v' &=& 3v(v-1)~.\label{eq:60a}
\end{eqnarray*}
The term $u'=0 \implies u=-\frac{\alpha_{1}}{2}$ further reduce the dynamical system as,
\begin{eqnarray}
 x' &=& \frac{x(y-3z-\frac{3\alpha_{1}}{2}+3)}{\alpha_{1}}~,\label{eq:56}\\
 y' &=& \frac{y(y-3z-\frac{5\alpha_{1}}{2}+3)}{\alpha_{1}}~,\label{eq:57}\\
 z' &=& -\left(\frac{3z+3\alpha_{1}-y-3}{2\alpha_{1}}(2z-3\alpha_{1})+\frac{3\kappa^{4}}{2}\frac{xv}{v-1}\right)~,\label{eq:58}\\
 v' &=& 3v(v-1)~.\label{eq:59}
\end{eqnarray}
In order to derive the dynamical features of the autonomous system, the coupled equation $x'=0$, $y'=0$, $z'=0$ and $v'=0$ are to be solved. TABLE-\ref{table:III} shows the corresponding critical points and their descriptions for the above system. In TABLE-\ref{table:IV}, stability conditions and the corresponding cosmological parameters are given. 
\begin{widetext}

\begin{table}[H]
\begin{center}
\renewcommand{\arraystretch}{2.5}
\caption{Critical points for the dynamical system.}\label{table:III}
\begin{tabular}{|c|c|c|c|c|c|}
\hline\hline
   Critical Points & $x$ & $y$ & $z$ & $v$ &  Exists For\\
   \hline\hline
   $A_2$ & $\gamma_{2}$ & $0$ & $\frac{3}{2}$ & $0$ &  $\alpha_{1}\neq -1$  \\\hline
   $B_2$ & $0$ & $0$ & $-\frac{3\alpha_{1}}{2}$ & $0$ &  $\alpha_{1}\neq 0$  \\\hline
   $C_2$ & $0$ & $0$ & $\frac{2+\alpha_{1}}{2}$ & $0$ &  $\alpha_{1}\neq 0$  \\\hline
   $D_2$ & $0$ & $-3-2\alpha_{1}$ & $-\frac{3\alpha_{1}}{2}$ & $0$ &  $\alpha_{1}\neq 0$  \\
   \hline
\end{tabular}
\end{center}
\end{table}

\begin{table}[H]
\begin{center}
\renewcommand{\arraystretch}{2.5}
\caption{Stability conditions, EoS parameter and deceleration parameter}\label{table:IV}
\begin{tabular}{|c|c|c|c|c|}
\hline\hline
   Critical Point & Stability Conditions  & $q$ & $\omega_{eff}$ & $\omega_{de}$\\
   \hline\hline
   $A_2$ & Unstable & $\frac{1}{2}$ & $0$ & $\frac{-7}{11}$  \\\hline
   $B_2$ & Unstable & $\frac{4\alpha_{1}+3}{2\alpha_{1}}$ & $\frac{\alpha_{1}+1}{\alpha_{1}2}$ & $\frac{\alpha_{1}(3\alpha_{1}-2)+2}{(4-7\alpha_{1})\alpha_{1}}$  \\\hline
   $C_2$ & \begin{tabular}{@{}c@{}} Stable for \\ $\alpha_{1} <-\frac{1}{2}\lor \alpha_{1} >0$ \end{tabular} & $-1$ & $-1$ & $-1$ \\\hline
   $D_2$ & Unstable & $1$ & $\frac{1}{3}$ & $\frac{4-13\alpha_{1}}{3(7\alpha_{1}-4)}$ \\
   \hline
\end{tabular}
\end{center}
\end{table}

\end{widetext}
FIG-- \ref{FIG--8} shows the phase portrait for the system given in Eqn. \eqref{eq:56}-\eqref{eq:59} with $\alpha_{1}=-1$ in $2-D$, $z$ vs $y$ plane. One can see that for $\alpha_{1}=-1$, we get stable point at $C_{2}(0,~0.5)$ and unstable point at $B_{2}(0,~1.5)$, $D_{2}(-1,~1.5)$. We have described in details the corresponding cosmology for each critical points as below:
\begin{itemize}
\item \textbf{Critical Point $A_{2}$ :} The point exists for $\alpha_{1}\neq -1$. The corresponding deceleration parameter and EoS parameter are $q=\frac{1}{2}$ and $\omega_{eff}=0$ respectively. The density parameters are $\Omega_{m}=\gamma_{2}$, $\Omega_{r}=0$ and $\Omega_{de}=0$ and if we consider $\gamma_{2}=1$ shows the matter dominated phase of the Universe. The eigenvalues of the Jacobian matrix for this critical point are given below. There is a zero, positive and negative signature of the eigenvalues, showing unstable saddle behavior.
\begin{equation*}
\{3,~-3,~-1,~0\}.
\end{equation*}
This unstable critical point can be seen in the phase portrait if drawn in the $xz$-plane.

\item \textbf{Critical Point $B_{2}$ :} The point exists for $\alpha_{1}\neq 0$. The corresponding deceleration parameter is $q=\frac{4\alpha_{1}+3}{2\alpha_{1}}$ and EoS parameter is $\frac{\alpha_{1}+1}{\alpha_{1}2}$. If we consider the $\alpha_{1}=-\frac{3}{2}$ then Universe shows radiation dominated phase and $\alpha_{1}=-1$ shows the matter dominated phase. The behaviour of the critical point leads to the decelerating phase of the Universe and the eigenvalues of Jacobian matrix for these critical points are zero, positive and negative signature, shows unstable saddle behavior.
 \begin{eqnarray*}
\left\{-3,~1,~4,~0\right\}~~~~~\left(\alpha_{1}=-\frac{3}{2}\right)~, \\
\left\{-3,~0,~3,~-3\right\}~~~~~\left(\alpha_{1}=-1\right).
\end{eqnarray*}

\item \textbf{Critical Point $C_{2}$ :} The points exists for $\alpha_{1}\neq 0$. The corresponding deceleration parameter and EoS parameter are $q=-1$ and $\omega_{eff}=-1$ respectively. The density parameters are $\Omega_{m}=0$, $\Omega_{r}=0$ and $\Omega_{de}=1$, i.e the Universe shows dark energy dominated phase. The eigenvalues of the Jacobian matrix for this critical point are given below. There is a negative signature of the eigenvalues, showing stable node behavior.
\begin{equation*}
\left\{-4,~-3,~-3,~-\frac{3 (2\alpha_{1}+1)}{\alpha_{1}}\right\}.
\end{equation*}

\item \textbf{Critical Point $D_{2}$ :} The points exists for $\alpha_{1}\neq 0$. The density parameters are $\Omega_{m}=0$, $\Omega_{r}=-3-2\alpha_{1}$ and $\Omega_{de}=0$. For $\alpha_{1}=-2$, the Universe shows the radiation dominated phase. The corresponding deceleration and EoS parameters are $q=1$ and $\omega_{eff}=\frac{1}{3}$ respectively. The eigenvalues for Jacobian matrix for corresponding critical points are positive and negative signature, thus shows unstable saddle behavior. Moreover the critical points leads to the decelerating phase of the Universe.

\begin{equation*}
\left\{-3,~1,~4,~-\frac{(2\alpha_{1}+3)}{\alpha_{1}}\right\}.
\end{equation*}
\end{itemize}

From TABLE- \ref{table:III}, we can see that the critical points are depends upon the model parameter $\alpha_{1}$. The model $\Phi(Q, T)= \alpha_{1}Q+\beta_{1}T^{2}-Q$, critical point $C_{2}$ showing the stable behavior under the condition on $\frac{4\alpha_{1}+3}{2\alpha_{1}}$. All other crititcal points shows the unstable beavior. $A_{2}$ shows the matter dominated phase of the Universe. Whereas critical point $B_{2}$ shows either matter dominated phase or radiation dominated phase of the Universe depending upon the value of $\alpha_{1}=-1$ or $\alpha_{1}=-\frac{3}{2}$ respectively. The critical point $D_{2}$ shows the radiation dominated phase of the Universe.

\section{Results and Discussions} \label{Sec.VI}
The dynamical parameters of $f(Q,T)$ gravity have been presented in the most general form for the function, $f(Q,T)= \alpha Q^m+\beta T$. Two cases pertaining to $m=1$ and $\beta =0$ are investigated for the late time behaviour of the Universe. The basic geometrical parameters such as the Hubble parameter and deceleration parameter are found to be in the preferred range of the cosmological observations and so also the EoS parameter. The geometrical parameters are scale factor dependent and hence depend on the value of $\Lambda$ only. Therefore since it is independent of model parameters, the present value of $H$ and $q$ does not change with the varying values of the model parameters.  Looking into the evolutionary behaviour, in both the cases the Universe may show the $\Lambda$CDM behaviour at late times. Though there is some difference in the present value of EoS and effective EoS parameter, but not very significant effect on the evolution. The key results and values of the parameters of both the cases are given in  Table \ref{table:V} and Table \ref{table:VI} below.
\begin{widetext}
    
\begin{table}[H]
\caption{Present value of Hubble parameter $H$, deceleration parameter $q$, EoS parameter $\omega$ and $\omega_{eff}.$
 }
\centering 
\begin{tabular}{c c c c c c c} 
\hline\hline 
\textbf{Model} &\quad \quad \textbf{Varying $\beta$} & \quad \quad \textbf{$H(z=0)$}  & \quad \quad \textbf{$q(z=0)$} & \quad \quad\textbf{$\omega(z=0)$} &\quad \quad \textbf{$\omega_{eff}(z=0)$} &  \\ [0.5ex] 
\hline
$\textbf{Case-I}$ 
& $\beta=0.01$ &  \quad \quad 71.808 & \quad \quad -0.701 & \quad \quad -0.803 & \quad \quad -0.866 &  \\
& $\beta=0.51$ &\quad \quad 71.808 & \quad \quad -0.701 & \quad \quad -0.797& \quad \quad -0.863 &\\
& $\beta=1.51$ &\quad \quad 71.808 & \quad \quad -0.701 & \quad \quad -0.791 & \quad \quad -0.860 &\\
\hline
\hline
$\textbf{Case-II}$ & \quad \quad\textbf{Varying $m$}&\\
\hline
&$m=0.8$& \quad \quad 71.808 & \quad \quad -0.701 & \quad \quad -0.838 & \quad \quad -0.799 &
 \\
& $m=1.0$& \quad \quad 71.808 & \quad \quad -0.701 & \quad \quad -0.799 & \quad \quad -0.799&
\\
& $m=1.2$ & \quad \quad 71.808 & \quad \quad -0.701 &  \quad \quad -0.761 & \quad \quad -0.799 &
 \\
\hline 
\end{tabular}
\label{table:V} 
\end{table}

\begin{table}[H]
\caption{Behaviour of energy conditions.}
\centering 
\begin{tabular}{c c c c c } 
\hline\hline 
\textbf{Model} &\quad  \textbf{Energy Conditions}&  \quad \textbf{Early Time}  & \quad \textbf{Present Time} &   \\ 
~~ &\quad  ~~&  \quad \textbf{($z>>0$)}  & \quad \textbf{($z=0$)} &  \\ [0.5ex] 
\hline
$\textbf{Case-I}$
 & SEC & Violated & Violated &  \\  & NEC & Satisfied & Satisfied &  \\  & DEC &  Satisfied & Satisfied &  \\
\hline
$\textbf{Case-II}$ 
 & SEC &  Violated & Violated & 
 \\  & NEC & Satisfied & Satisfied & \\ & DEC &  Satisfied & Satisfied &  \\
\hline 
\end{tabular}
\label{table:VI} 
\end{table}
\end{widetext}

We have derived the dynamical system analysis for the general $f(Q,T)$ model by considering two forms of $\Phi(Q,T)$ in order to find the stability behaviour of the models. The corresponding critical points are obtained and the eigenvalues are derived.  Model-I corresponds to the case $m=1$ with general model. One can see the stability for $m=1$ in TABLE- \ref{table:II}. Model-II corresponds to $\beta=0$ in the general model. Similar to model I, the stability of model II is mentioned in TABLE- \ref{table:II} with the corresponding critical points. From the TABLE \ref{table:I}-\ref{table:II}, we can conclude that both the models are stable.  The TABLE- \ref{table:III} and TABLE-\ref{table:IV} provide the detail behaviour of the critical points and the stability condition of the model. Finally, we conclude that we have presented the stable accelerating cosmological models framed in extended symmetric teleparallel gravity.

\section*{Acknowledgement}
LP acknowledges Department of Science and Technology (DST), Govt. of India, New Delhi for awarding INSPIRE fellowship (File No. DST/INSPIRE Fellowship/2019/IF190600) to carry out the research work as Senior Research Fellow. SKT and BM  acknowledge IUCAA, Pune, India for providing support through the visiting Associateship program. The authors are thankful to the anonymous reviewer for the comments and suggestions to improve the quality of the manuscript.

\end{document}